\documentclass{aa}

\usepackage{graphicx}
\usepackage{txfonts}
\usepackage{hyperref}

\DeclareGraphicsExtensions{.eps}

\begin{document}

\title{Dimming events of evolved stars due to clouds of molecular gas}
\subtitle{Scenarios based on 3D radiation-hydrodynamics simulations with CO5BOLD}

\titlerunning{Dimming of evolved stars} 

\author{
  Bernd Freytag\inst{1}
    \and
  Susanne H{\"o}fner\inst{1}
    \and
  Bernhard Aringer\inst{2}
    \and
  Andrea Chiavassa\inst{3}
}

\institute{
  Theoretical Astrophysics,
  Department of Physics and Astronomy,
  Uppsala University,
  Box~516,
  SE-751~20 Uppsala,
  Sweden \\
  \email{Bernd.Freytag@physics.uu.se}
    \and
  Department of Astrophysics,
  University of Vienna,
  T{\"u}rkenschanzstrasse~17,
  A-1180 Vienna,
  Austria
    \and
  Université C{\^o}te d'Azur,
  Observatoire de la C{\^o}te d'Azur,
  CNRS,
  Lagrange,
  CS~34229, Nice,
  France
}

\date{\today}

\abstract
{
The dramatic dimming episode of the red supergiant Betelgeuse in 2019 and 2020,
caused by a partial darkening of the stellar disk,
has highlighted gaps in the understanding of the evolution of massive stars.
}
{
We analyzed numerical models
to investigate the processes behind the formation of dark surface patches
and the associated reduction in the disk-integrated stellar light.
}
{
With the CO5BOLD code,
we performed global 3D radiation-hydrodynamical simulations of evolved stars,
including convection in the stellar interior,
self-excited pulsations,
and the resulting atmospheric dynamics with strong radiative shocks.
}
{
We attribute dimming phenomena to obscuring clouds of cool gas in the lower atmosphere,
forming according to three different scenarios.
One process transports material outward in a strong shock,
similar to what occurs in 1D simulations of radially pulsating
asymptotic giant branch (AGB) stars.
However, in 3D models, deviations from spherical symmetry of the shock front
can lead to further local density enhancements.
Another mechanism is triggered by a large convective upflow structure,
in combination with exceptionally strong radial pulsations.
This induces Rayleigh-Taylor instabilities,
causing plumes of material to be sent outward into the atmosphere.
The third and rarest scenario involves large-amplitude convective fluctuations,
leading to enhanced flows in deep downdrafts,
which rebound and send material outward.
In all cases, the dense gas above the stellar surface cools and darkens rapidly
in visible light.
AGB stars show  localized dark patches regularly
during intermediate phases of their large-amplitude pulsations,
while more massive stars will only intermittently form such patches
during luminosity minima.
}
{
The episodic levitation of dense gas clumps above the stellar surface,
followed by the formation of complex molecules in the cooling gas
and possibly dust grains at a later stage,
can account for the dark patches and strong dimming events
of supergiant stars such as Betelgeuse.
}


\maketitle

\section{Introduction}\label{s:Intro}

Some red supergiants (RSGs) are among the brightest stars in the night sky,
which has turned them into popular objects for amateur astronomers,
who have made significant contributions
to the study of the long-term variability of the stellar light curves.
The large apparent diameter of the stars, on the other hand,
renders them favored targets for professional interferometric observations,
revealing dynamical large-scale surface structures.

Early interferometric images of the RSG Betelgeuse ($\alpha$~Ori)
with WHT and COAST
\citep{Buscher1990MNRAS.245P...7B,
Wilson1992MNRAS.257..369W,
Burns1997MNRAS.290L..11B,
Tuthill1997MNRAS.285..529T,
Wilson1997MNRAS.291..819W,
Young2000MNRAS.315..635Y},
were reconstructed using models with one or more hot spots on the stellar disk.
However, \cite{Young2000MNRAS.315..635Y} achieved an equally good fit to the
interferometric data with a model of a disk and a dark spot.
Together, the observations demonstrated the dynamical evolution
of large-scale surface structures, which
in most cases were attributed to giant convection cells
\citep{Schwarzschild1975ApJ...195..137S}.
\cite{Gray2000ApJ...532..487G} came to a similar conclusion based on
the analysis of spectra.

The Great Dimming of Betelgeuse during the winter of 2019-2020
\citep{Guinan2019ATel13341....1G},
which was visible to the naked eye
\citep[see, for example, the analyses of (B)AAVSO data in][]{Joyce2020ApJ...902...63J,
Lloyd2020BAAVC.184...22L},
triggered an intense observational effort
\citep[e.g.,][]{Kashyap2020ATel13501....1K,
Gehrz2020ATel13518....1G,
Sbordone2020ATel13525....1S,
Dupree2020ApJ...899...68D,
Dharmawardena2020ApJ...897L...9D,
Harper2020ApJ...893L..23H,
Harper2020ApJ...905...34H,
Mittag2023A&A...669A...9M,
Cannon2023A&A...675A..46C}.
The proximity of the star has made it possible to obtain resolved images of
its surface, showing that only parts of the star appeared to be much darker
(\citealt{Montarges2021Natur.594..365M};
for more details, the reader is referred to the recent review of
\citealt{Wheeler2023A&G....64.3.11W}).

\begin{table*}[htb]
 \begin{center}
  \caption{
    \label{t:ModelParam}
     Basic model parameters and derived quantities 
     }
  \begin{tabular}{lrrrrrrrrrrl}
\hline\hline
model & $M_\star$ & $M_\mathrm{env}$ & $L_\star$ & $n_x^3$ & $x_\mathrm{box}$ & $t_\mathrm{avg}$ & $R_{\star,s_\mathrm{min}}$ & $T_{\mathrm{eff},s_\mathrm{min}}$ & $\log g_{s_\mathrm{min}}$ & $P_\mathrm{puls}$ & \\
 & $(M_\sun)$ & $(M_\star)$ & $(L_\sun)$ &  & $(R_\sun)$ & (yr) & $(R_\sun)$ & (K) & cgs & (d) & \\ \hline
st28gm06n051 &  1.0 & 0.182 &  6933 & 637$^3$ & 1558 & 48.27 &  328 & 2907 & -0.598 &  459 & \\
st28gm05n056 &  1.0 & 0.399 &  5085 & 317$^3$ & 1263 & 60.20 &  293 & 2848 & -0.498 &  390 & shock, Sect.\,\ref{s:ShockDarkening} \\
st28gm05n032 &  1.5 & 0.256 &  6936 & 637$^3$ & 1263 & 24.93 &  280 & 3149 & -0.286 &  256 & \\
st35gm04n048 &  5.0 & 0.110 & 40914 & 637$^3$ & 1929 & 26.58 &  623 & 3288 & -0.456 &  523 & rebound, Sect.\,\ref{s:NonstationaryConvection} \\
st35gm04n045 &  5.0 & 0.092 & 41336 & 637$^3$ & 1929 & 20.91 &  597 & 3368 & -0.419 &  516 & plume, Sect.\,\ref{s:Plumes} \\
st34gm02n002 &  8.0 & 0.277 & 41675 & 765$^3$ & 1626 & 23.17 &  602 & 3359 & -0.223 &  314   \\
\hline
  \end{tabular}
 \end{center}
{
  The table shows the model name
  {(composed of
   “s,” referring to the gravitational potential,
   ``t'' for ``temperature'' followed by two digits indicating the approximate
     effective temperature,
   ``g'' for ``gravity'' followed by ``m'' for ``minus'' and two digits indicating
     the approximate logarithm of the surface gravity,
   and ``n'' for ``number'' with the model counter)};
  the current stellar mass, $M_\star$, used for prescribing the gravitational potential;
  the envelope mass, $M_\mathrm{env}$, derived from integrating the mass density
  of all grid cells within the computational box; 
  the average emitted luminosity, $L_\star$; 
  the grid dimensions, $n_x^3$;
  the edge length of the cubical computational box, $x_\mathrm{box}$;
  the time, $t_\mathrm{avg}$, used for averaging the remaining quantities in this table;
  the average approximate stellar radius, $R_{\star,s_\mathrm{min}}$,
  taken as the radial position of the innermost photospheric entropy minimum
  \citep[see][]{Ahmad2023A&A...669A..49A};
  the average approximate effective temperature, $T_{\mathrm{eff},s_\mathrm{min}}$;
  the logarithm of the average approximate surface gravity, $\log g_{s_\mathrm{min}}$;
  the pulsation period, $P_\mathrm{puls}$;
  and a comment about the levitation mechanism that we used the model as a reference for.
  The first three models (with $M_\star$\,=\,1 and 1.5\,$M_\sun$) are examples of the scenario
  of darkening due to cool post-shock gas
  (see Sect.\,\ref{s:ShockDarkening}).
  The 5\,$M_\sun$ model st35gm04n048 shows a prominent dimming event
  due to a convective rebound
  (we present in Sect.\,\ref{s:NonstationaryConvection}).
  The remaining two models (with $M_\star$\,=\,5 and 8\,$M_\sun$) show dark patches due to plumes
  (see Sect.\,\ref{s:Plumes}).
}
\end{table*}

Surprisingly little is known about the final phases of the lives of massive stars.
To understand the physics behind the observations, they need to be
interpreted with self-consistent models, based on first principles.
We used the CO5BOLD code
\citep{Freytag2012JCP...231..919F,
Freytag2017A&A...600A.137F},
which allows for global 3D simulations of radiation-hydrodynamical (RHD) processes
in stars on the asymptotic giant branch (AGB) and RSGs,
including large-scale convection in the stellar interior,
self-excited pulsations,
and the resulting atmospheric dynamics with strong radiative shock waves. 
In contrast to RSGs, the less massive AGB stars tend to show strong periodic
luminosity variations caused by large-amplitude pulsations.
A well-known example is Mira (o~Cet),
which was one of the first variable stars discovered.
The global AGB-star models presented by \cite{Freytag2017A&A...600A.137F} show
self-excited radial pulsations, with periods that are in good agreement
with observations of Mira variables \citep[see][]{Ahmad2023A&A...669A..49A}
and realistic convective surface structures
\citep[see the discussion in][]{Paladini2018Natur.553..310P}.\footnote{Some
animations are available under
\href{https://www.astro.uu.se/~bf/movie/AGBmovie.html}{''3D models of AGB stars''}}.
A complex inhomogeneous distribution of atmospheric gas emerges naturally in these models,
as a consequence of large-scale convective flows below the photosphere
and the resulting network of atmospheric shock waves.
As is shown in detail by \cite{Hoefner2019A&A...623A.158H},
the dynamical patterns in the gas are imprinted on the dust
in the close stellar environment,
due to the density- and temperature-sensitivity of the grain growth process,
explaining the origin of clumpy dust clouds observed around AGB stars
\citep[e.g.,][]{Ohnaka2016A&A...589A..91O,
Ohnaka2017A&A...597A..20O,
Khouri2016A&A...591A..70K,
Kervella2018A&A...609A..67K,
Khouri2018A&A...620A..75K}.
Global 3D models of AGB stars and their dust-driven winds,
recently presented by \cite{Freytag2023A&A...669A.155F},
show that the inhomogeneities persist in the wind-acceleration region,
leading to clumpy outflows.

In red (super)giant stars,
the characteristic convective pattern of hot upflow regions,
surrounded by cool lanes with downflowing material,
is mostly hidden below the visible layers.
However, based on a selection of global 3D RHD simulations
produced with CO5BOLD,
we demonstrate how subsurface convection
in conjunction with acoustic pulsations
can strongly affect the structure and appearance
of the convectively stable photosphere.
We showcase in the current paper
three mechanisms for the formation of cool, dense clumps of gas
in the lower stellar atmosphere
that result in dark patches when seen against the stellar surface.
We discuss how these processes may produce strong visual dimming events,
with little effect on the shape of disk-integrated spectra.
In the end, we suggest that convection and pulsations,
which are widely believed to be the responsible drivers for the
Great Dimming of Betelgeuse
\citep[see the comprehensive overview by][]{Wheeler2023A&G....64.3.11W},
can achieve this via
shock waves and high-Mach-number overshooting motions.

In Sect.\,\ref{s:Setup}, we describe the general model setup and
show examples of the formation of dark patches
found in various models of cool giant stars,
following one of three scenarios that can lead to strong local dimming events.
We discuss the first one in detail in Sect.\,\ref{s:Plumes}
and show the influence on stellar spectra in Sect.\,\ref{s:Spectra}.
The second process, which is able to produce comparably rare strong dimming events,
is presented in Sect.\,\ref{s:NonstationaryConvection}.
The third one is a modification of a well-known mechanism found in dynamical
1D models and is summarized in Sect.\,\ref{s:ShockDarkening}.
In Sect.\,\ref{s:Discussion}, the new results are discussed and briefly
confronted with observations,
while the conclusions are given in Sect.\,\ref{s:Conclusions}.

\section{Model setup and overview of simulations}\label{s:Setup}

\subsection{General setup}\label{s:GeneralSetup}

In our global ``star-in-a-box'' simulations of evolved stars,
the CO5BOLD code
\citep{Freytag2012JCP...231..919F,
Freytag2013MSAIS..24...26F,
Freytag2017MmSAI..88...12F}
numerically integrates the coupled nonlinear equations of
compressible hydrodynamics with an approximate Roe solver
and nonlocal radiative energy transfer
with a short-characteristics scheme.
The grid is Cartesian;
the computational domain and all grid cells are cubical.

The fixed gravitational potential is spherically symmetric.
In the outer layers, the radial profile corresponds to a given point-like core mass,
but it is smoothed in the central region of the star 
\citep[see Eq.\,(41) in][]{Freytag2012JCP...231..919F}.  
The comparably tiny central nuclear-reaction region cannot
be resolved with grid cells of constant size.
Instead, in the core region
(with a typical radial extension of around 20\,\% of the stellar radius),
a source term feeds in energy.
{A drag force is active only in this small volume
to limit the contribution of this region to the driving of global dipolar convection flows,
without suppressing such flows completely.}
All outer boundaries are open for radiation and the flow of matter.

\begin{figure*}[hbtp]
\begin{center}
\scalebox{1.0}{\includegraphics[width=6cm]{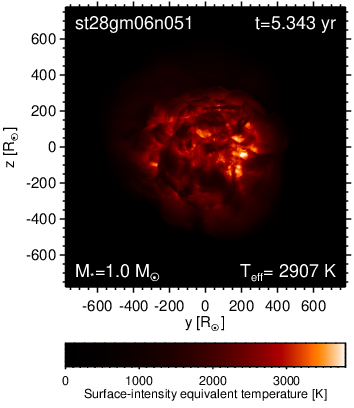}}
\scalebox{1.0}{\includegraphics[width=6cm]{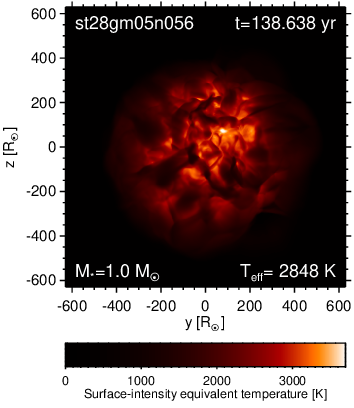}}
\scalebox{1.0}{\includegraphics[width=6cm]{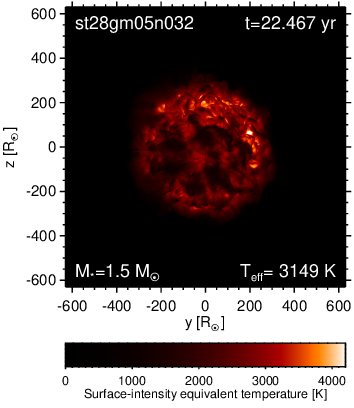}}

\scalebox{1.0}{\includegraphics[width=6cm]{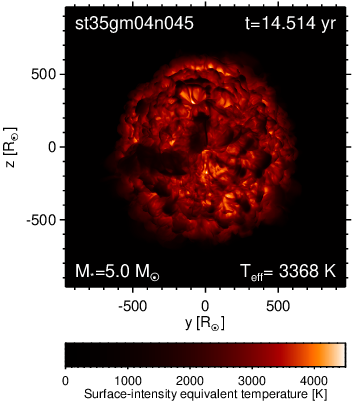}}
\scalebox{1.0}{\includegraphics[width=6cm]{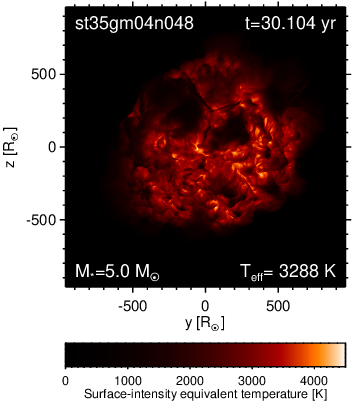}}
\scalebox{1.0}{\includegraphics[width=6cm]{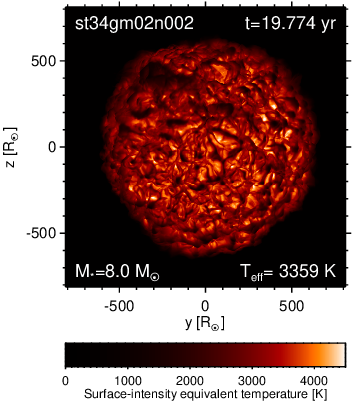}}
\end{center}
\caption{
  \label{f:stXX_Int}
  Examples of dark patches forming in models with different stellar parameters.
  Shown are bolometric intensity maps
  of the models in Tab.\,\ref{t:ModelParam}.
  Top row (AGB stars):
  st28gm06n051 ($M_\star$\,=\,1\,$M_\sun$, $T_\mathrm{eff}$\,$\approx$\,2907\,K),
  st28gm06n056 ($M_\star$\,=\,1\,$M_\sun$, $T_\mathrm{eff}$\,$\approx$\,2848\,K;
                opposite side of model shown in
                Fig.\,\ref{f:st28gm05n056_0873500_QuSeq1}),
  st28gm05n032 ($M_\star$\,=\,1.5\,$M_\sun$, $T_\mathrm{eff}$\,$\approx$\,3149\,K).
  Bottom row (RSG stars):
  st35gm04n045 ($M_\star$\,=\,5\,$M_\sun$, $T_\mathrm{eff}$\,$\approx$\,3368\,K;
                different snapshot of our standard model shown in
                Fig.\,\ref{f:st35gm04n045_0069300_QuSeq1}),
  st35gm04n048 ($M_\star$\,=\,5\,$M_\sun$, $T_\mathrm{eff}$\,$\approx$\,3288\,K;
                opposite side of model shown in
                Fig.\,\ref{f:st35gm04n048_0232501_QuSeq1}),
  st34gm02n002 ($M_\star$\,=\,8\,$M_\sun$, $T_\mathrm{eff}$\,$\approx$\,3359\,K).
}
\end{figure*}

\begin{figure*}[hbtp]
\begin{center}
\scalebox{1.0}{\includegraphics[width=18cm]{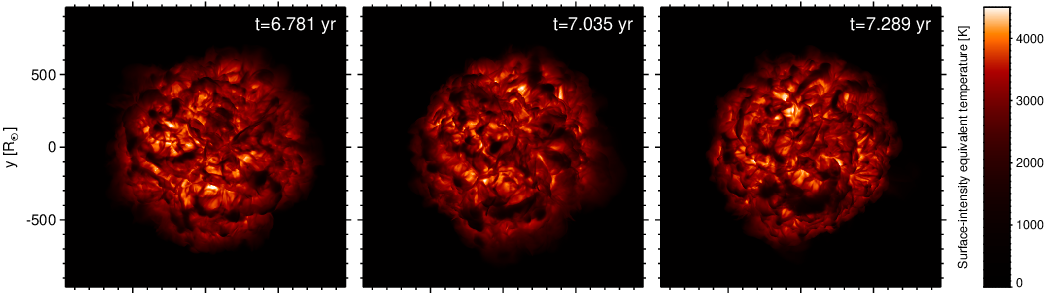}}

\scalebox{1.0}{\includegraphics[width=18cm]{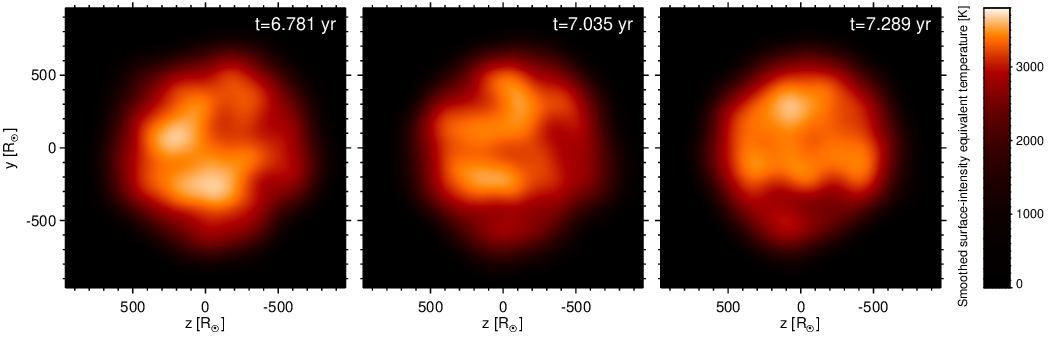}}
\end{center}
\caption{
  \label{f:st35gm04n045_0069300_QuSeq1}
  Time sequence of
  bolometric intensity
  of three snapshots of the 5\,$M_\sun$ model st35gm04n045
  in original resolution (top row)
  and blurred by a Gaussian ($\sigma$\,=\,30\,px\,$\approx$\,$91\,R_\sun$, bottom row)
  with an adjusted color scale.
  We focus our analysis on the dark curved horizontal patch
  developing on the lower hemisphere of the stellar disc
  {(the dark feature centered around $y$\,=\,-400\,R$_\sun$, $z$\,=\,0
  in the top right frame).}
}
\end{figure*}

\begin{figure*}[hbtp]
\begin{center}
\scalebox{1.0}{\includegraphics[width=18cm]{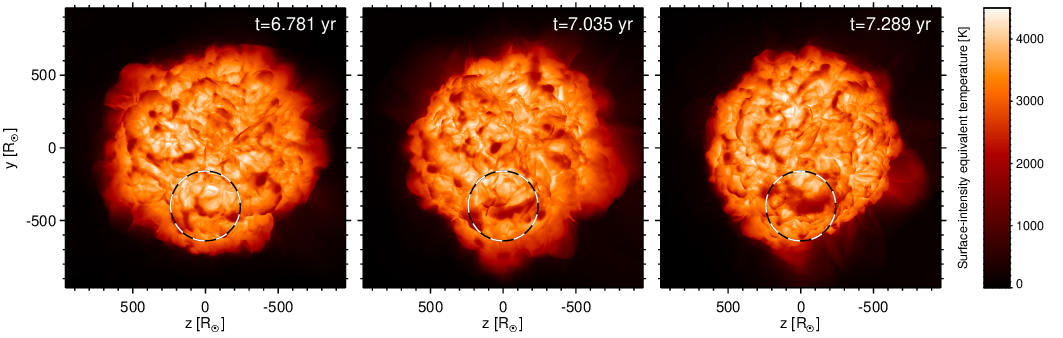}}\vspace{-0.1cm}

\scalebox{1.0}{\includegraphics[width=18cm]{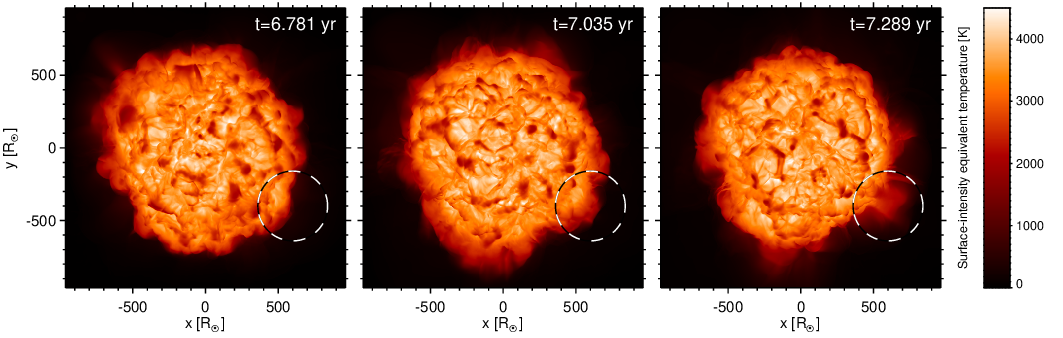}}\vspace{-0.1cm}

\scalebox{1.0}{\includegraphics[width=18cm]{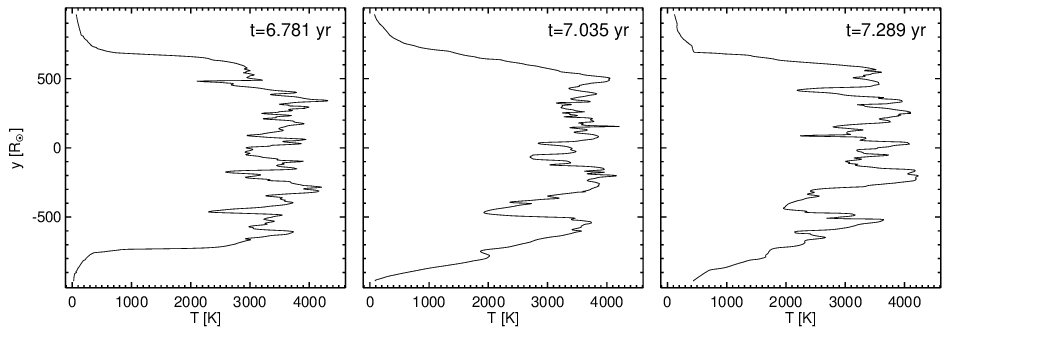}}\vspace{-0.1cm}
\end{center}
\caption{
  \label{f:st35gm04n045_0069300_QuSeq2}
  Time sequences of
  surface temperature data derived from the bolometric intensity
  of three snapshots of the 5\,$M_\sun$ model st35gm04n045.
  Top and center row:
  Temperature maps
  for two viewing directions.
  The geometrical axes differ between the rows.
  The top row corresponds to Fig.\,\ref{f:st35gm04n045_0069300_QuSeq1}.
  The rotated view in the bottom row shows the object in the top row
  seen by an observer on the left.
  The cool material causing the dark horizontal patch
  developing on the lower hemisphere in the top row
  (at about $x$\,=\,600\,R$_\sun$, $y$\,=\,-400\,R$_\sun$, $z$\,=\,0)
  is seen from the side at 4~o'clock in the bottom row as a narrower emitting feature,
  marked by a dashed-line circle.
  It corresponds to the cross sections
  in Fig.\,\ref{f:st35gm04n045_0069300_QuSeq3}.
  Bottom row:
  Central vertical ($z$\,=\,0; $x$ is the viewing direction) cuts
  through the temperature maps
  in the top row.
  The dark horizontal patch developing in that sequence
  produces the extended temperature minimum
  with values around $T$\,=\,2000\,K
  at about $y$\,=\,-400\,R$_\sun$, here.
}
\end{figure*}

\begin{figure*}[hbtp]
\begin{center}
\scalebox{1.0}{\includegraphics[width=18cm]{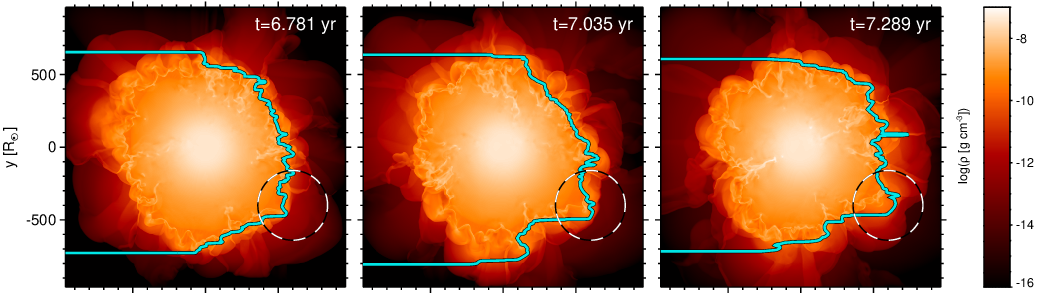}}

\scalebox{1.0}{\includegraphics[width=18cm]{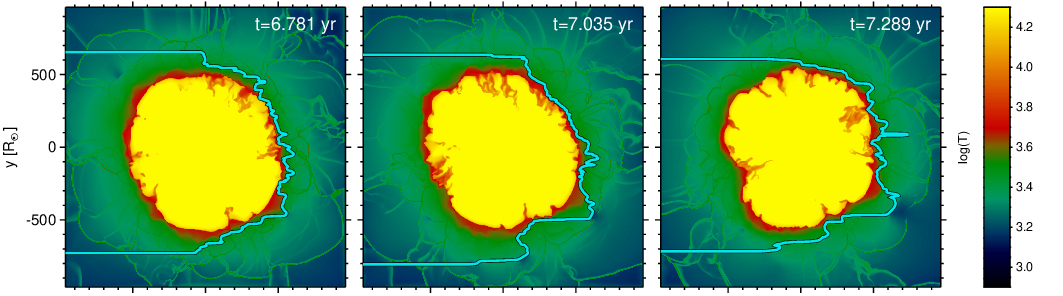}}

\scalebox{1.0}{\includegraphics[width=18cm]{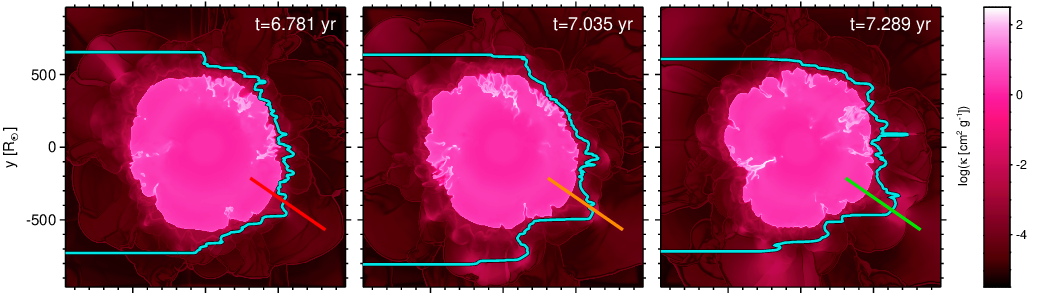}}

\scalebox{1.0}{\includegraphics[width=18cm]{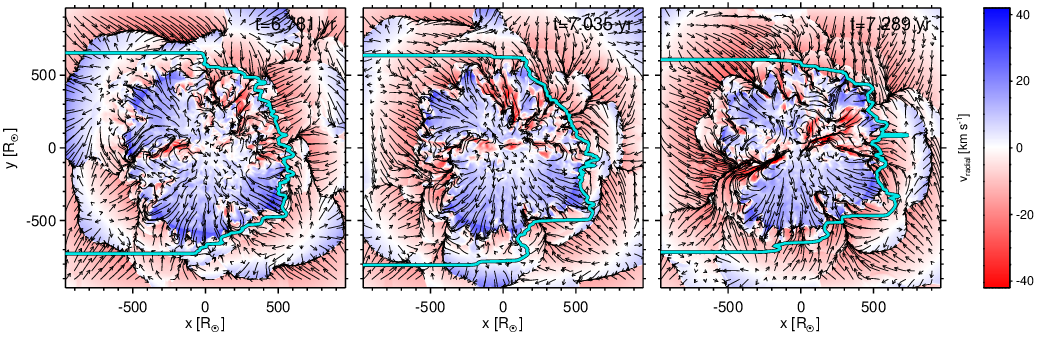}}
\end{center}
\caption{
  \label{f:st35gm04n045_0069300_QuSeq3}
  Time sequences of cross sections through the center of the 5\,$M_\sun$~model st35gm04n045
  showing
  the logarithm of density,
  the logarithm of temperature
    (scale restricted to values below $T$=20\,000\,K
    to better display the conditions within the atmosphere),
  the logarithm of opacity (per gram)
    with colored lines (red, orange, green)
    indicating the spatial positions of the radial profiles
    in Fig.\,\ref{f:st35gm04n045_1DRay},
  and the radial velocity
    with overlaid pseudo-streamlines.
  The view (axes $x$ and $y$) corresponds to the one in the middle row
    of Fig.\,\ref{f:st35gm04n045_0069300_QuSeq2}
    with the dashed-line circle marking the main region of interest.
  The cyan contour line indicates gray optical depth unity ($\tau_\mathrm{Ross}$\,=\,1)
  for an observer located to the right.
  The cool, dense material causing the dark patch or emission region
  in Fig.\,\ref{f:st35gm04n045_0069300_QuSeq1}
  is seen from the side at 4~o'clock,
  causing the extension of the $\tau_\mathrm{Ross}$ contour line to the right.\vspace*{1cm}
}
\end{figure*}

\begin{figure*}[hbtp]
\begin{center}
\scalebox{1.0}{\includegraphics[width=18cm]{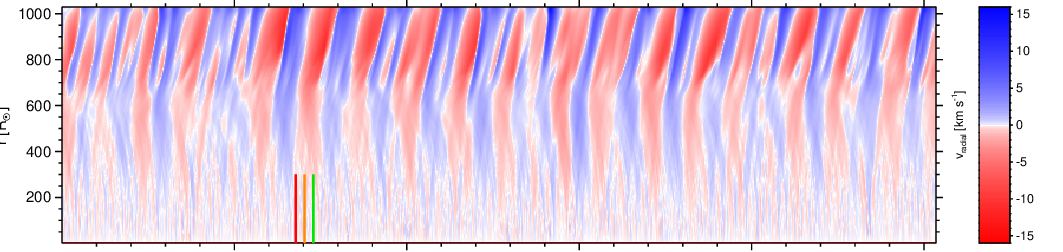}}
\scalebox{1.0}{\includegraphics[width=16.0cm]{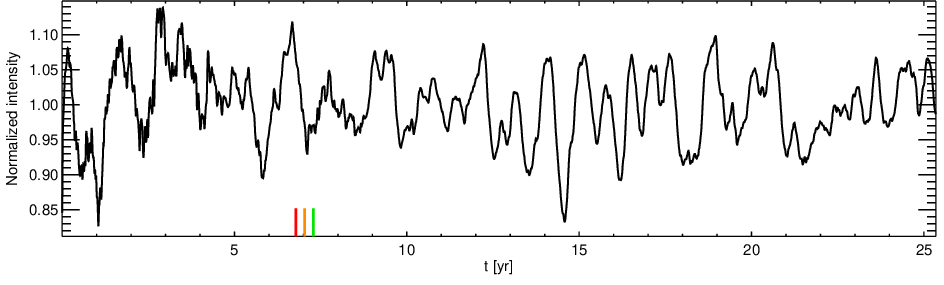}\hspace*{2cm}}
\end{center}
\caption{
  \label{f:st35gm04n045_QuOvertimeAndx}
  Temporal behavior of radial velocity and intensity
  of the 5\,$M_\sun$ model st35gm04n045.
  Top panel: Spherical averages of the
  radial velocity
  as a function of radial distance and time.
  Bottom panel: Normalized intensity (luminosity seen by an observer)
  corresponding to the view in Figs.\,\ref{f:st35gm04n045_0069300_QuSeq1} and \ref{f:st35gm04n045_0069300_QuSeq2}.
  The colored vertical lines (red, orange, green) at about $t$\,=\,7\,yr
  mark the instants selected for
  Figs.\,\ref{f:st35gm04n045_0069300_QuSeq1} to \ref{f:spectra_BA}.
}
\end{figure*}

\begin{figure*}[ht]
\begin{center}
\includegraphics[width=6.0cm]{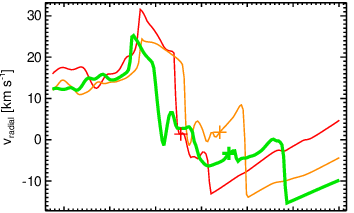}
\includegraphics[width=6.0cm]{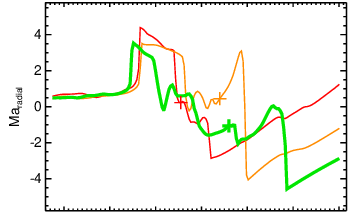}
\includegraphics[width=6.0cm]{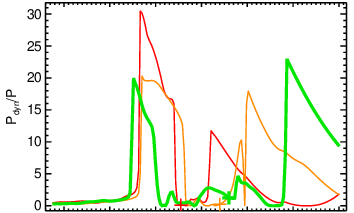}

\includegraphics[width=6.0cm]{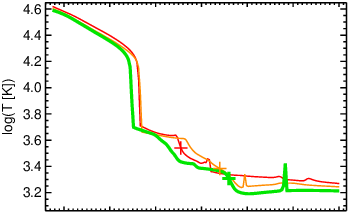}
\includegraphics[width=6.0cm]{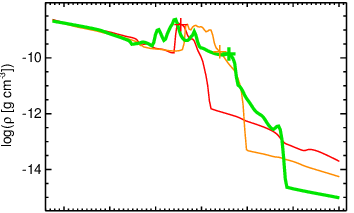}
\includegraphics[width=6.0cm]{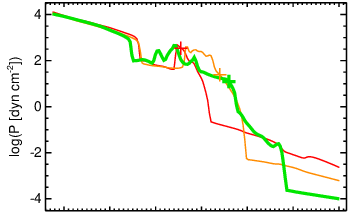}

\includegraphics[width=6.0cm]{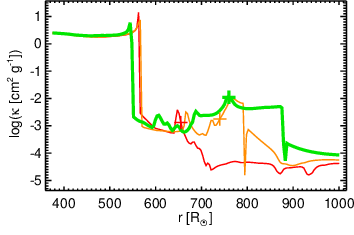}
\includegraphics[width=6.0cm]{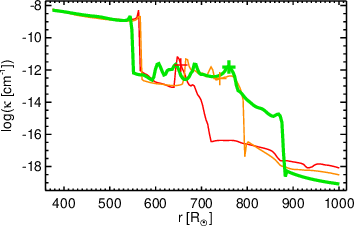}
\includegraphics[width=6.0cm]{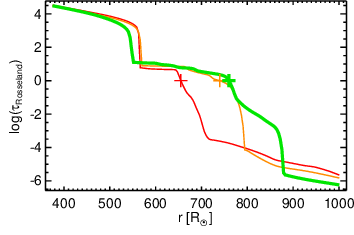}\vspace*{-0.1cm}
\end{center}
\caption{
  \label{f:st35gm04n045_1DRay}
  One-dimensional radial profiles of the 5\,$M_\sun$ model st35gm04n045
  showing
  radial velocity $v_\mathrm{radial}$,
  {radial Mach number $\mathrm{Ma}_\mathrm{radial}$\,$=$\,$v_\mathrm{radial}/c_\mathrm{sound}$},
  dynamical pressure due to radial velocity over gas pressure
    $\rho v_\mathrm{radial}^2/P$,
  logarithm of temperature $\log(T)$,
  logarithm of gas density $\log(\rho)$,
  logarithm of pressure $\log(P)$,
  logarithm of Rosseland opacity per mass unit $\log(\kappa [\mathrm{cm}^2 \mathrm{g}^{-1}])$,
  logarithm of Rosseland opacity per volume $\log(\kappa [\mathrm{cm}^{-1}])$, and
  logarithm of Rosseland optical depth along the ray $\log(\tau_\mathrm{Ross})$.
  The colors (red, orange, green) indicate the three time steps selected in
  Figs.\,\ref{f:st35gm04n045_0069300_QuSeq1} to \ref{f:st35gm04n045_0069300_QuSeq3},
  about 8$\cdot$10$^6$\,s (92\,d) apart.
  The plus signs mark the position of optical depth unity ($\tau_\mathrm{Ross}$\,=\,1),
  integrated along the ray.
}
\end{figure*}

\begin{figure*}[hbtp]

\begin{center}
\scalebox{1.0}{\includegraphics[width=6.00cm]{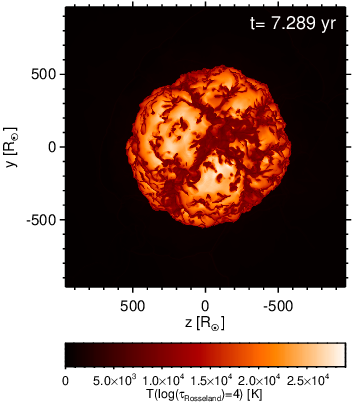}}
\scalebox{1.0}{\includegraphics[width=5.04cm]{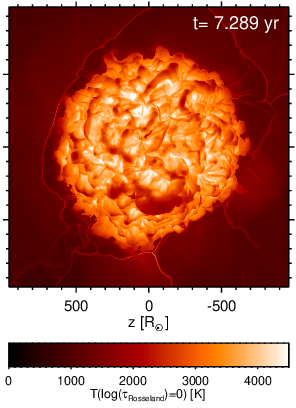}}
\scalebox{1.0}{\includegraphics[width=5.04cm]{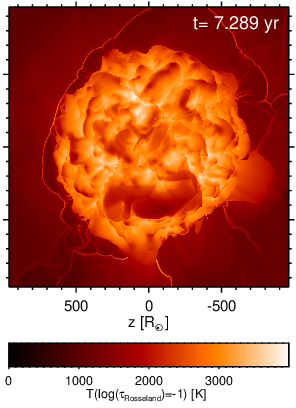}}

\scalebox{1.0}{\includegraphics[width=6.00cm]{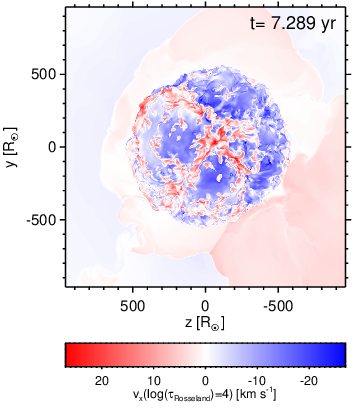}}
\scalebox{1.0}{\includegraphics[width=5.04cm]{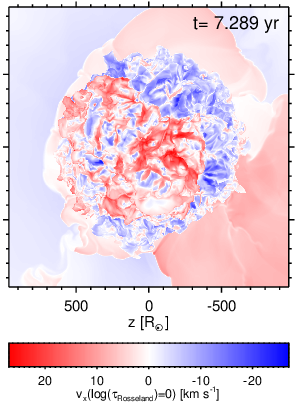}}
\scalebox{1.0}{\includegraphics[width=5.04cm]{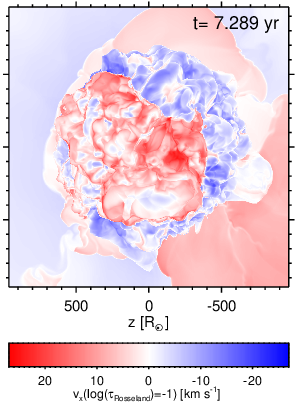}}\vspace{-0.1cm}
\end{center}
\caption{
  \label{f:st35gm04n045_0072609_QtauSeq}
  Temperatures (top row)
  and line-of-sight velocities (bottom row, blue indicates motion toward the observer)
  at optical depths $\log(\tau_\mathrm{Ross})$\,=\,4, 0, and -1 (from left to right)
  for the 5\,$M_\sun$ model st35gm04n045.
  At points with an optical depths lower than the nominal value,
  the value at the rear boundary is plotted.
  The signature of the cool material causing the dark horizontal patch
  developing on the lower hemisphere
  is visible around
  $y$\,=\,-400\,$R_\sun$, $z$\,=\,0 at $\log(\tau_\mathrm{Ross})$\,=\,0 and -1.
  The view (axes $y$ and $z$) corresponds to the intensity maps
  in Fig.\,\ref{f:st35gm04n045_0069300_QuSeq1}.
}
\end{figure*}

{The tables for atmospheric gas opacities used here 
are based on COMA data \citep[see][]{Aringer2000DissAri, Aringer2016MNRAS.457.3611A}.
Abundances assumed are
  (X/Y/Z)=(0.7088/0.2750/0.0162).
We followed the assumptions in \cite{Aringer2016MNRAS.457.3611A},
based on
\cite{Caffau2009MmSAI..80..643C,
Caffau2009A&A...498..877C}.
The resulting carbon to oxygen ratio is close to 0.55.
The low-temperature COMA table
is extended with OPAL data \citep[see][]{Iglesias1992ApJ...397..717I}
at temperatures above approximately 12\,000\,K.
Scattering is treated as true absorption.
The tabulated equation of state takes into account the ionization of hydrogen
and helium, the formation of H$_2$ molecules, and a neutral metal.
The assumed abundances
  (X/Y/Z)=(0.7035/0.2793/0.0172)
differ slightly from the values for the COMA tables,
most likely without any noticeable consequences.}
Dust and radiation pressure are not taken into account in the
models discussed in this paper.

\subsection{Specific model parameters}

An overview of the models used in this paper, and of their parameters, is given
in Tab.\,\ref{t:ModelParam}.
While the initial conditions
for the first global AGB star simulations
used in \cite{Freytag2008A&A...483..571F}
were based on hydrostatic 1D stratifications,
all later AGB models were derived from predecessor 3D models.
In the current paper, we analyze in detail
a new high-resolution model of a 5\,$M_\sun$ red-giant star
(model st35gm04n045 with 637$^3$ grid points in Tab.\,\ref{t:ModelParam}).
It was obtained from an older model (st35gm04n37 with 315$^3$ points)
by interpolating to a finer grid
and adding a few grid layers at the boundaries to be able to follow shocks further out.
Model st35gm04n048 was derived from model st35gm04n045 by gently increasing the
mass contained in the numerical box ($M_\mathrm{env}$ in Tab.\,\ref{t:ModelParam})
at the beginning of the simulations.
To demonstrate how the overall dynamical properties, appearances of dark patches, 
or global dimming events
depend on stellar parameters, we also briefly discuss several 
other simulations
listed in \,Tab.\,\ref{t:ModelParam}.
A more detailed description of the model-setup procedure
and, for instance, of checks for thermal and dynamical relaxation
to skip initial phases with transient phenomena,
is given in \cite{Freytag2017A&A...600A.137F}.

\subsection{Examples of local surface-darkening events}\label{s:SurfaceDarkeningOverview}

In Fig.\,\ref{f:stXX_Int}, snapshots of the bolometric surface intensity
are shown for models with a range of masses
(see Tab.\,\ref{t:ModelParam}).
The plots were derived directly from the data produced
during the simulations.
The color table was chosen to be appropriate for a cool red star
but is otherwise arbitrary.
{We transformed the intensity units to more intuitive
equivalent radiation temperature values,
but otherwise in plots with identical coloring to what we used before
\citep[e.g., in][]{Freytag2012JCP...231..919F,
Freytag2017A&A...600A.137F,
Freytag2023A&A...669A.155F}.}

We manually selected instances at which large dark areas occur.
The darkening in the 1 and 1.5\,$M_\sun$ AGB models can be almost global,
possibly with some local brightenings due to inhomogeneities,
induced while the shock front passes through the lower atmosphere.
The instances in Fig.\,\ref{f:stXX_Int} (top row) were selected
to depict large-scale but not global dimming events,
which are therefore not extreme but very common phenomena
at {stages between minima and maxima} in the pulsation cycle
(see Sect.\,\ref{s:ShockDarkening}).

The models with 5 and 8\,$M_\sun$ have smaller granules
relative to the stellar diameters
due to the larger radii.
Pulsation amplitudes are smaller and effective temperatures are higher
than in the lower-mass models,
preventing global darkening phases.
Accordingly, the snapshots were selected
such that large-scale -- comparably rare -- dark patches are visible
(see Sects.\,\ref{s:Plumes} and \ref{s:NonstationaryConvection}).

In all models, the large dark areas are never uniform but show substructures.
Although the patches seem rather similar at first glance,
we find three different mechanisms
contributing to their production to varying degrees,
depending on the stellar parameters.
These will be discussed in the following section.

\section{Plumes due to Rayleigh-Taylor instabilities}\label{s:Plumes}

We start our discussion of dimming mechanisms with a scenario
that frequently occurs in our simulations of AGB and RSG stars.
It involves the emission of puffs of material
above the outer boundary of the convection zone,
which cool and lead to dark areas on the stellar surface.
Under certain circumstances,
a more powerful plume might levitate sufficient material
to produce an extended patch,
causing a noticeable dimming of the disk-integrated stellar light,
as is discussed below.

\subsection{{Overview of analysis methods}}

For our analysis,
we selected three snapshots of the 5\,$M_\sun$ model st35gm04n045
(see Tab.\,\ref{t:ModelParam}),
at instants about 3~months apart,
featuring a dark and simple-shaped patch at a location on the surface
that makes plotting easy.
Otherwise, the patch is in no way special
(neither the largest nor smallest, for example).

The development of the dark patch is documented by means of time sequences
of maps of the emitted bolometric intensity
in Fig.\,\ref{f:st35gm04n045_0069300_QuSeq1}.
Figure\,\ref{f:st35gm04n045_0069300_QuSeq2} (top and middle row)
shows maps of the derived surface temperature
{(on this plot with the linear temperature chosen for the color scale)}
seen from different viewing angles toward the star.
The bottom row of Fig.\,\ref{f:st35gm04n045_0069300_QuSeq2}
shows spatial profiles (1D stripes at $z$=0)
of the surface-temperature distribution
that is displayed in the maps in the top row.
We concentrate on the dark horizontally elongated patch
prominent in the lower hemisphere
of the images of Fig.\,\ref{f:st35gm04n045_0069300_QuSeq1}.
The mechanism behind the development of the patch is best deduced from the
time sequences of central slices in
Fig.\,\ref{f:st35gm04n045_0069300_QuSeq3}.

To investigate purely radial motions (for instance, radial pulsations), 
we used averages over spherical shells
of the radial mass flux $\langle \rho v_\mathrm{radial} \rangle_\Omega(r,t)$
and the mass density $\langle \rho \rangle_\Omega(r,t)$ for each snapshot, 
computed during the CO5BOLD run.
In a post-processing step,
we took the ratio of these quantities as the radial velocity
and assembled the data from all time steps into one data field,
which is now a function,
$\langle v_\mathrm{radial} \rangle(r,t)$,
of the radius coordinate
and time, plotted in Fig.\,\ref{f:st35gm04n045_QuOvertimeAndx}. 
The colored bars (red, orange, and green) in the plot 
indicate the three instants used in the other figures.
A detailed analysis of pulsations from this type of data for a number of models is presented in
\cite{Ahmad2023A&A...669A..49A}.

Numerical values of several quantities
along the selected radial ray through the patch
{(marked in the third row of panels in Fig.\,\ref{f:st35gm04n045_0069300_QuSeq3})}
can be read off Fig.\,\ref{f:st35gm04n045_1DRay}.
Figure\,\ref{f:st35gm04n045_0072609_QtauSeq}
shows the distributions of
temperature and line-of-sight velocity
at $\log(\tau_\mathrm{Ross})$\,=\,4,0,-1
for the selected instant
to visualize the correlation between the deep and the photospheric layers.
The time evolution of the opacities along the ray
is put into context in Fig.\,\ref{f:st35gm04n045_opa}.

\subsection{General model properties}\label{s:GeneralModel}

The dynamical behavior of the selected 5\,$M_\sun$ model
is summarized in the following.
It is governed by similar processes to those in cooler 1\,$M_\sun$ models
of AGB stars described in detail in \cite{Freytag2017A&A...600A.137F}.

The energy inserted in the central region of the model at a constant rate
is transported outward mostly by large-scale convective cells,
with downdrafts occasionally reaching from the stellar surface
all the way into the core region.
These huge cells hardly show up directly in maps of the bolometric intensity
(Fig.\,\ref{f:st35gm04n045_0069300_QuSeq1})
or of the derived surface temperature
(Fig.\,\ref{f:st35gm04n045_0069300_QuSeq2}).
Still, the central cuts through the model in Fig.\,\ref{f:st35gm04n045_0069300_QuSeq3}
as well as the temperature and velocity distributions at $\log(\tau_\mathrm{Ross})$\,=\,4
(Fig.\,\ref{f:st35gm04n045_0072609_QtauSeq})
clearly show the large scale of the interior structures.
For example, the rightmost snapshot of the velocity field
in Fig.\,\ref{f:st35gm04n045_0069300_QuSeq3}
shows only two dominating downdrafts on opposite sides of the star.

The small-scale low-contrast granular pattern
with bright upflows and darker surrounding intergranular lanes,
known from the solar surface,
is here and there (with typical scales of several 10\,$R_\sun$)
visible in Figs.\,\ref{f:st35gm04n045_0069300_QuSeq1}
and \ref{f:st35gm04n045_0069300_QuSeq2}.
These granules,
as well as the irregularities at and below the surface of the convection zone
at $r$\,$\approx$\,550\,$R_\sun$ in all quantities
displayed in Fig.\,\ref{f:st35gm04n045_0069300_QuSeq3}, 
are caused by surface convection cells much smaller than the interior cells.
However, the granules are most often hidden
under prominent dark patches of various sizes
(Fig.\,\ref{f:st35gm04n045_0069300_QuSeq1}),
which are due to cooler material above the surface of the convection zone.
This is discussed in detail in Sect.\,\ref{s:DimmingScenario}.

The stellar surface is characterized by a sharp, steep inward increase in opacities
caused by the partial ionization of hydrogen
(see Fig.\,\ref{f:st35gm04n045_0069300_QuSeq3}
and also the radial opacity profiles in Fig.\,\ref{f:st35gm04n045_1DRay}
at $r$\,$\approx$\,550\,$R_\sun$).
This is responsible for the sharp jump in temperature from photospheric conditions
to values above $T$=10\,000\,K
(the transition from green over red to yellow in the temperature slices
in the second row of Fig.\,\ref{f:st35gm04n045_0069300_QuSeq3}
and the strong rise in the temperature profiles in Fig.\,\ref{f:st35gm04n045_1DRay}).
{The entropy shows a drastic inward increase as well,
which marks the top of the convectively unstable zone
\citep[see, e.g., Fig.\,2 in][]{Freytag2017A&A...600A.137F}.}
Under close to hydrostatic conditions,
the relative pressure increase would be small across the temperature jump,
causing a drop in gas density as the temperature rises inward.
The associated density inversion is the main driver of convection
and is still visible in the dynamical models
as a bright orange rugged line above the slightly darker orange regions
in the density plots in the top panels of Fig.\,\ref{f:st35gm04n045_0069300_QuSeq3}
and local maxima between $r$\,=\,600 and 720\,$R_\sun$
in the density profiles in Fig.\,\ref{f:st35gm04n045_1DRay}.
{In this case.
the temperature step and the density-inversion layers are detached
due to the action of the dynamical pressure.
This is discussed in detail in Sect.\,\ref{s:DimmingScenario}.}

The model shows slightly irregular acoustic pulsations in the fundamental radial mode,
which are clearly visible in the time evolution of the averaged radial velocity
and the light curve
in Fig.\,\ref{f:st35gm04n045_QuOvertimeAndx}.
The period is about 1.4~years.
The predominantly standing-wave (i.e., vertical) pattern
between $r$\,$\approx$\,300\,$R_\sun$ and $r$\,$\approx$\,650\,$R_\sun$
is only weakly recognizable in the innermost regions ($\lessapprox$\,200\,$R_\sun$)
due to the signature of fluctuations in the net radial mass flow
caused by chaotic convective motions.
The dominant radial pulsations are accompanied by a dipolar mode and
small-scale traveling waves.
The latter are generated as noise of the nonstationary convective flow
\citep[see][]{Nordlund2001ApJ...546..576N,
Stein2001ApJ...546..585S}.
These oscillations and waves have small amplitudes in the interior and
are not directly visible in any of the presented cross-section plots.

However, in the cooler outer atmospheric layers
with a lower density and lower sound speed,
the waves turn into predominantly outward-traveling shocks,
which govern the local dynamics
\citep[resulting in inclined streaks in Fig.\,\ref{f:st35gm04n045_QuOvertimeAndx};
see also][]{Freytag2017A&A...600A.137F, Ahmad2023A&A...669A..49A}.
The shock waves are clearly visible, for example, in density
(top row of Fig.\,\ref{f:st35gm04n045_0069300_QuSeq3})
and velocity slices
(bottom row of Fig.\,\ref{f:st35gm04n045_0069300_QuSeq3})
and the corresponding profiles in Fig.\,\ref{f:st35gm04n045_1DRay}.
The dissipation of energy in the shock fronts leads to a rise in temperature,
apparent as a pattern of thin green lines on a blue background in the temperature slices
in the second row of panels in Fig.\,\ref{f:st35gm04n045_0069300_QuSeq3}
and as spikes in the temperature profiles in Fig.\,\ref{f:st35gm04n045_1DRay}.
This will be discussed further in Sect.\,\ref{s:ChromoSphere}.

\subsection{Formation of a dark patch}\label{s:DimmingScenario}

In red (super)giant stars,
both nonstationary convective flows and pulsations
have relatively high Mach numbers compared to the Sun
\citep[see the plot of Mach numbers vs.\ pressure for a set of CO5BOLD models
  from the main sequence to supergiant stars in Fig.\,6 of][]{Freytag2013EAS....60..137F}.

The Mach number, Ma, along the selected rays
in Fig.\,\ref{f:st35gm04n045_1DRay}
shows that at the top of the convection zone
the Mach numbers are around one,
while they are significantly larger in parts of the thin atmosphere above,
and thus emerging structures should be interpreted
in terms of hydrodynamic processes,
in contrast to a pressure-equilibrium situation.
Transonic velocities and strong density contrasts
lead to a highly unstable situation and nonstationary flows.
The pressure profiles are far from monotonic.

Conservation of mass requires that
convective upflows slow considerably
when entering the higher-density surface layers,
resulting in a strong gradient of the velocity and the dynamical pressure.
%
%
In regions with sufficiently large radial velocities
(due to a combination of 
a strong convection flow and a small-scale wave or
global radial pulsation{, see Sect.\,\ref{s:PulsationsConvection}}),
the high-density surface layers are lifted up by the dynamical pressure.
%
In addition, Rayleigh-Taylor instabilities develop,
occurring when a high-velocity low-density flow of matter
encounters a layer of denser material,
causing material to overshoot the boundary of the normal convection zone
in puffs of cool gas.
This is visible in the density, temperature, and velocity slices
in Fig.\,\ref{f:st35gm04n045_0069300_QuSeq3},
in particular in the region 
marked by the colored radial lines in the opacity slices,
which indicate where the data plotted in Fig.\,\ref{f:st35gm04n045_1DRay}
have been extracted from.
The plume structure with an inward zone of material moving toward the observer
and an outer region of matter falling back
is apparent in the lower right panel
of Fig.\,\ref{f:st35gm04n045_0072609_QtauSeq}
The increase in density no longer coincides with the drop in temperature,
as is expected in a pressure-equilibrium situation.
Instead, the atmospheric high-density material is lifted upward
by the dynamical pressure,
which is enormous compared to the gas pressure in the region below,
(see Fig.\,\ref{f:st35gm04n045_1DRay}).
The density-inversion layer
forms a zigzag pattern around the star
just inside the cool atmosphere, where the density drops rapidly outward,
(see the rugged orange line at around $r$\,$\approx$\,600\,$R_\sun$ with
extrusions inward and outward of about 100\,$R_\sun$\
in the top row of panels in Fig.\,\ref{f:st35gm04n045_0069300_QuSeq3}).
This might be a more vigorous version
of the large cusp-like features at the stellar surface
in high-resolution 2D RHD models of a Cepheid
found by \cite{Mundprecht2013MNRAS.435.3191M}.

The wave that has aided in pushing up the material
transforms into a shock and travels rapidly further out.
It is clearly visible in the middle and right panel of the bottom-row slices
in Fig.\,\ref{f:st35gm04n045_0069300_QuSeq3}
and in many quantities in Fig.\,\ref{f:st35gm04n045_1DRay}.
The event occurs at a time when the average atmospheric mass flow
is directed inward (see Fig.\,\ref{f:st35gm04n045_QuOvertimeAndx}),
in contrast to the outward motion of the dense gas puff.

Since the mechanism for pushing up the material
requires the combined effects of an outward-traveling wave
and a convective upflow (that cannot comprise the entire stellar surface),
the resulting patch may be small or large but cannot be global.
This is in contrast to the situation in 
models of cooler, less massive stars
with lower surface gravities and stronger radial pulsations,
in which an acoustic shock wave alone is sufficient to lift up the gas
(see Sect.\,\ref{s:ShockDarkening}).

Over time, the temperature of the elevated material drops due to
adiabatic expansion in a convectively stable environment and
radiative cooling.
The upper part of the elevated gas radiates energy away more efficiently
than a normal stellar surface element,
because the spatial angle from which it receives radiation is smaller,
while the angle into which it radiates more energy than it receives is larger.
That leads to a further cooling down
(see the formation of the blue blob in the temperature slices
in Fig.\,\ref{f:st35gm04n045_0069300_QuSeq3}, second row).
In addition, the optically thick lower part of the elevated matter shields
the upper part from direct light from the hot stellar surface.

\begin{figure}[hbtp]
\begin{center}
\includegraphics[width=7.3cm]{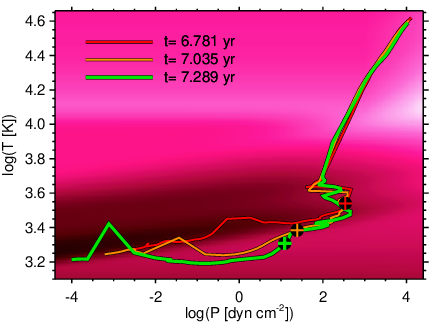}\includegraphics[width=1.5cm]{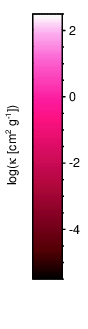}
\end{center}
\caption{
  \label{f:st35gm04n045_opa}
  {Logarithm of} Rosseland opacity per mass unit [cm$^2$\,g$^{-1}$]
  as a function of the logarithm of pressure
  and logarithm of temperature
  with overlaid 1D profiles from Fig.\,\ref{f:st35gm04n045_1DRay}.
  The color table is the same as for the opacity-per-gram panel
  in Fig.\,\ref{f:st35gm04n045_0069300_QuSeq3}.
  The plus signs mark the position of optical depth unity ($\tau_\mathrm{Ross}$\,=\,1),
  integrated along the ray.
}
\end{figure}

When the temperature reaches values below 2000\,K, depending on pressure,
opacities increase strongly due to the formation of molecules
(see the build-up of the magenta blob in the third row of opacity slices
in Fig.\,\ref{f:st35gm04n045_0069300_QuSeq3},
the radial opacity profiles in Fig.\,\ref{f:st35gm04n045_1DRay},
and Fig.\,\ref{f:st35gm04n045_opa}).
That leads to further shielding of the upper layers of the elevated material
from the stellar radiation,
more efficient emission, and
therefore an even stronger cooling (see Sect.\,\ref{s:GasOpacities} for a more detailed discussion of the main
contributors to the opacities).

An extended (but not global) dark patch has formed,
causing local darkening,
which is even visible after taking instrumental blurring crudely into account
(compare the two rows of original and blurred intensity maps
in Fig.\,\ref{f:st35gm04n045_0069300_QuSeq1}).
%
A sufficiently large patch will lead to a dimming of the surface-integrated
stellar light.
%
As the temperature of the dark patch
and the corresponding amount of emitted light is 
low compared to the rest of the visible stellar surface
(see Fig.\,\ref{f:st35gm04n045_0069300_QuSeq2}),
the change in the optical stellar spectrum may be small.

\subsection{The role of pulsations and convection for the plumes}\label{s:PulsationsConvection}

The maps of the temperature and the line-of-sight velocity
(Fig.\,\ref{f:st35gm04n045_0072609_QtauSeq})
at three optical depths
show the complex combined effect of large, deep convection cells,
smaller surface granules, acoustic waves, and shocks.
In the deeper layers, dominating cool downdrafts
(roughly forming an “X” in the top right panel),
separate just a few global upflow regions,
which are obvious in the temperature and the velocity maps.
Prominent features such as
the strong downdraft just above the center of the image
(the center of the “X”)
and the expanding upflow region in the upper right hemisphere of the star
imprint their signatures on the atmosphere.
There, small-scale granules and oscillatory motions render the
situation much more complex.

We emphasize that the predominant net line-of-sight atmospheric velocities toward the
observer in the upper right part of the star in
Fig.\,\ref{f:st35gm04n045_0072609_QtauSeq} and the prevailing recessing velocities in the center and lower
right part are not a sign of rotation,
but due to the underlying nonstationary convective motions.
\cite{Kravchenko2019A&A...632A..28K}
show similar large-scale surface motions in their Fig.\,12 and
\cite{Ma2024ApJ...962L..36M}
analyze the consequences for measurements of rotation rates.

It is not straightforward to apply standard considerations about
the stability or instability of a buoyant blob of gas,
because these rely on pressure equilibrium between the blob and the surrounding matter.
This is not fulfilled in the environment of a molecular puff.
Still, it reaches into heights
well above the region of convective instability according to standard criteria.
And while the kinetic energy flux is directed inward in the convection zone
(due to the narrow downdrafts in comparison to the wide upflow regions),
it carries about 2\% of the total stellar luminosity outward close to the surface
(due to the focused outward flow in the high-velocity puffs).

Such an event occurs in the wake of an outward-traveling shock wave,
well behind the shock front,
with a lifetime longer than the travel time of the shock
through the lower atmosphere.
Similar but weaker structures,
which can exist without the push by a shock wave,
are visible all over the surface
(see Figs.\,\ref{f:st35gm04n045_0069300_QuSeq1}
and \ref{f:st35gm04n045_0069300_QuSeq2}).
The cause of a particularly large plume could be
an enhancement in the large-scale convective flow
and/or a temporal increase in the pulsation amplitude
and/or a superposition of, for instance, the fundamental radial
and dipole mode.

\begin{figure}[hbtp]
\begin{center}
\includegraphics[width=8.8cm]{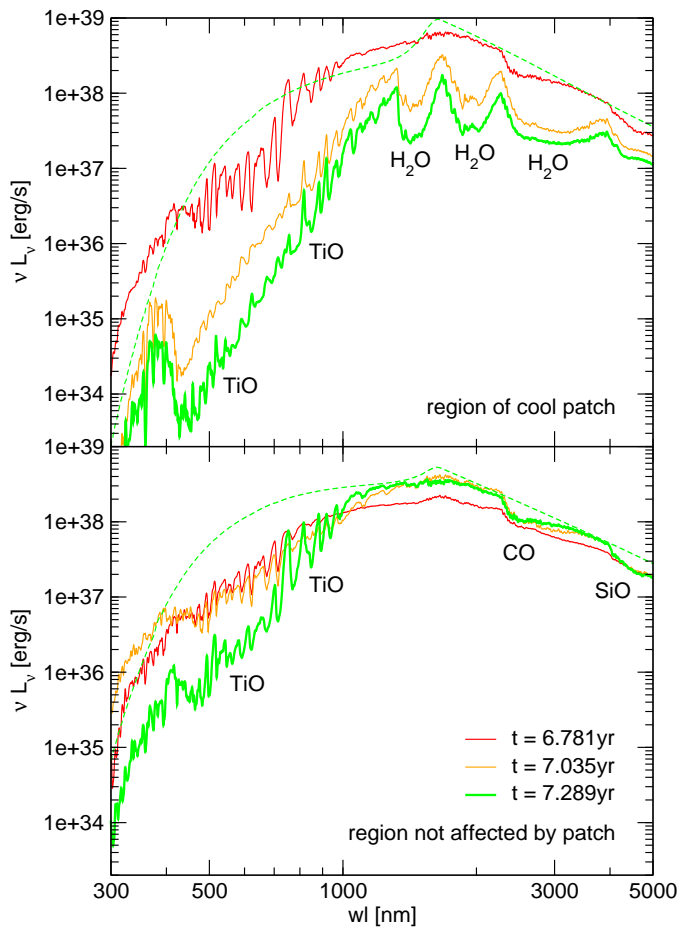}
\end{center}
\caption{
  \label{f:spectra_BA}
  Low-resolution ($R$=200) 1D spectra based on the 5\,$M_\sun$ model st35gm04n045.
  The results shown were computed along rays from the stellar center
  passing through a certain spot on the surface.
  The energy distributions in the upper panel
  correspond to the radial structures and time steps
  displayed in Fig.\,\ref{f:st35gm04n045_1DRay}
  (rays in the third row of Fig.\,\ref{f:st35gm04n045_0069300_QuSeq3}),
  which represent the region of the dark patch.
  The spectra in the lower panel covering the same epochs are typical for
  areas not affected by the local cooling. They were computed for horizontal
  rays in Fig.\,\ref{f:st35gm04n045_0069300_QuSeq3}.
  All displayed intensities correspond to an expansion of the
  fluxes to a uniform disk with the size of the star. The dashed line represents
  the continuum obtained by neglecting all atomic and molecular transitions for
  the last time step.
  The most important molecular features have been indicated.
}
\end{figure}

\subsection{Effects of molecular opacities and stellar spectra}\label{s:GasOpacities}\label{s:Spectra}

As was mentioned above,
a strong increase in the gas opacities toward low temperatures ($\lesssim$\,2000\,K)
plays an important role in the cooling of the gas and the appearance of dark patches.
This is illustrated in Fig.\,\ref{f:st35gm04n045_opa},
showing that $\kappa_\mathrm{Ross}$
has a minimum in the interval between approximately 2000\,K and 4000\,K,
depending on gas pressure.
The opacity rise above about 4000\,K occurs
because of larger continuous absorption due to the ionization of hydrogen.

The very pronounced and sudden increase in the opacity
when the temperature drops below 1500 to 2500\,K,
depending on the gas pressure,
is caused by millions of overlapping molecular lines.
The transition zone gets warmer with higher pressure,
which can be explained by the fact that larger densities
favor the formation of molecules.
For the assumed solar-like chemical composition, the most important contributors
to the opacity are TiO in the optical range below about 1\,$\mu$m and H$_2$O in
the near- and mid-infrared. Other species included in the calculations, which may
have some influence on the resulting $\kappa_\mathrm{Ross}$, are CO, CH, CN,
SiO, OH, VO, FeH, and CrH\@. More information concerning the COMA opacity data used
can be found in \cite{Aringer2016MNRAS.457.3611A}.

The atmospheric radial profile at $\tau_\mathrm{Ross}$\,$<$\,1
selected for the first time step, $t$\,=\,6.781\,yr
(without a major dark patch),
remains mostly inside the low-opacity range
(red curve in Fig.\,\ref{f:st35gm04n045_opa}).
Following the peak in temperature due to the shock traveling
into regions with lower pressure
(to the left in Fig.\,\ref{f:st35gm04n045_opa}),
the dip in temperature caused by the elevated material
deepens and widens with time.
The atmospheric radial profile intersecting the dark patch at $t$\,=\,7.289\,yr
(green curve in Fig.\,\ref{f:st35gm04n045_opa})
passes through the large-opacity region
dominated by intense molecular absorption
(at $\log T$\,$\approx$\,3.2,
between $\log P$\,$\approx$\,-2.3 and $\log P$\,$\approx$\,0.7),
shifting the continuum-formation layer at $\tau_\mathrm{Ross}$\,=\,1
further out into regions with lower temperatures
(see also Fig.\,\ref{f:st35gm04n045_1DRay}).
The molecules form in intermittent cool and dense puffs close to the star.
The structures are dynamical and complex,
deviating from the spherical-symmetric layer above the photosphere
assumed by \cite{Tsuji2000ApJ...540L..99T}.

In Fig.\,\ref{f:spectra_BA},
we compare the temporal variation in the spectral energy distribution
emitted by two different locations on the stellar surface in model st35gm04n045.
The three time steps discussed in the previous sections
(see Figs.\,\ref{f:st35gm04n045_0069300_QuSeq1} to \ref{f:st35gm04n045_opa})
are covered.
In the upper panel, we show spectra from the area where a dark patch develops,
while the spectra displayed in the lower one represent a part of the surface
not affected by this local cooling. The plotted data correspond
to an observer placed directly over those regions (central ray).
All relevant continuous and atomic or molecular line
opacities have been included, assuming solar abundances
and a microturbulent velocity of 2.5\,km/s.
More details concerning the methods of the spectrum synthesis
can be found in
\cite{Aringer2016MNRAS.457.3611A,
Aringer2019MNRAS.487.2133A}.
The presented spectra have a resolution of R=200
and cover the range from 300 to 5000\,nm (visual and near-infrared).
Effects of velocity fields were not taken into account.

\begin{figure*}[ht]
\begin{center}
\scalebox{1.0}{\includegraphics[width=18.0cm]{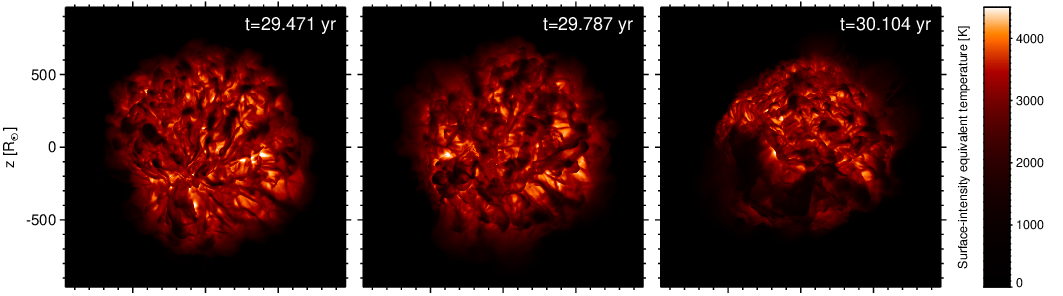}}

\scalebox{1.0}{\includegraphics[width=18.0cm]{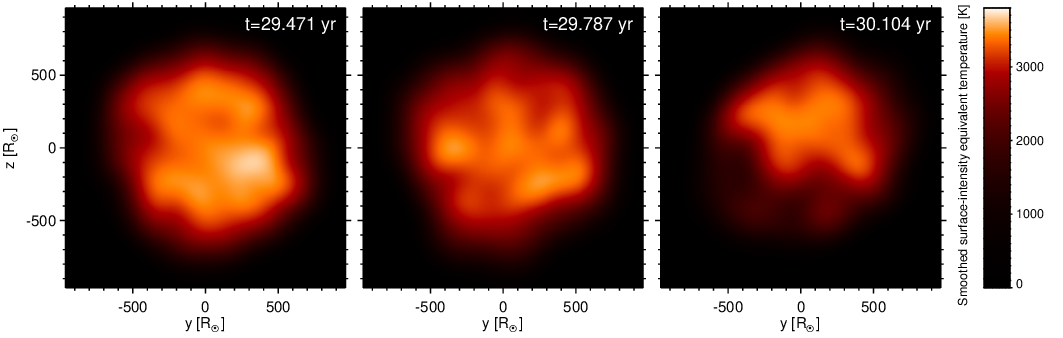}}
\end{center}
\caption{
  \label{f:st35gm04n048_0232501_QuSeq1}
  Time sequence of
  bolometric intensity
  of three snapshots of the 5\,$M_\sun$ model st35gm04n048
  in original resolution (top row)
  and blurred by a Gaussian ($\sigma$\,=\,30\,px\,$\approx$\,$91\,R_\sun$, bottom row,
  with an adjusted color scale).
}
\end{figure*}

\begin{figure*}[hbtp]
\begin{center}
\scalebox{1.0}{\includegraphics[width=18.0cm]{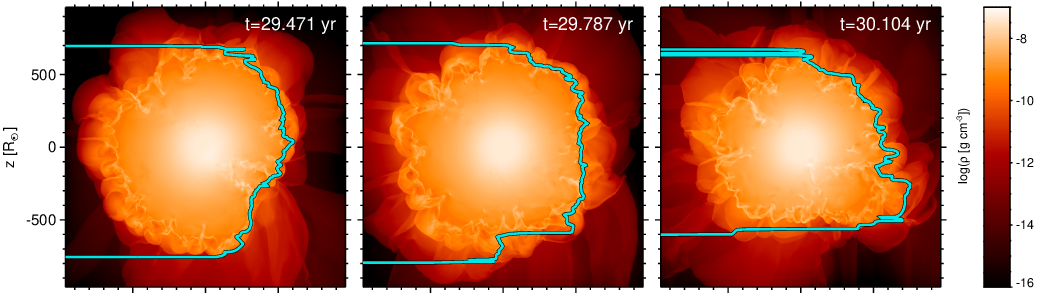}}

\scalebox{1.0}{\includegraphics[width=18.0cm]{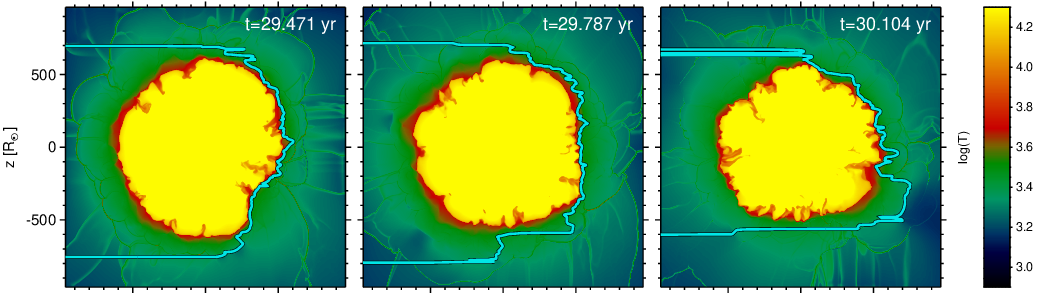}}

\scalebox{1.0}{\includegraphics[width=18.0cm]{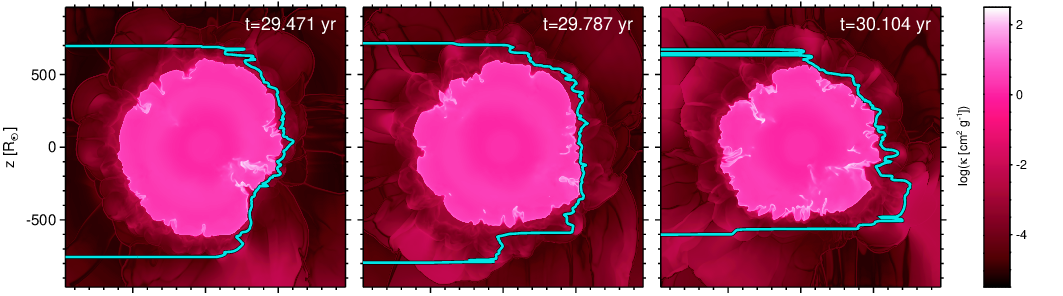}}

\scalebox{1.0}{\includegraphics[width=18.0cm]{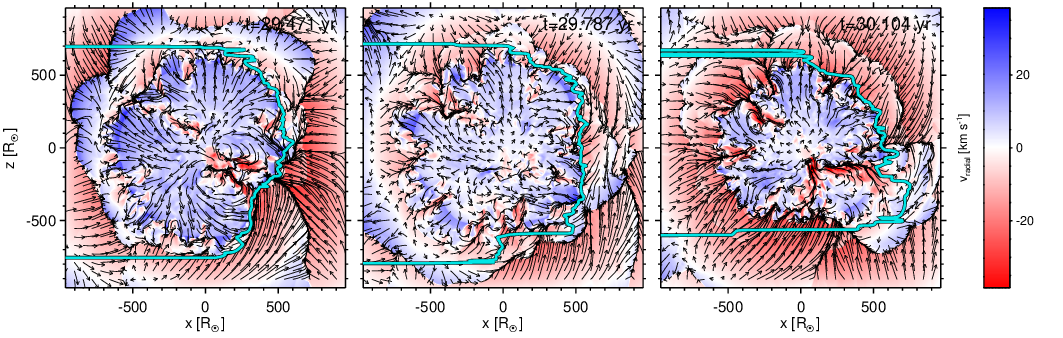}}
\end{center}
\caption{
  \label{f:st35gm04n048_0232501_QuSeq3}
  Time sequences of cross sections through the center of the 5\,$M_\sun$ model st35gm04n048
  showing
  the logarithm of density,
  the logarithm of temperature
    (restricted to values below $T$\,=\,20\,000\,K),
  the logarithm of opacity (per gram),
  and the radial velocity
    with overlaid pseudo-streamlines.
  The cyan contour line indicates optical depth unity ($\tau_\mathrm{Ross}$\,=\,1)
  for an observer {on} the right,
  corresponding to the viewing direction shown in Fig.\,\ref{f:st35gm04n048_0232501_QuSeq1}.\vspace*{1.5cm}
}
\end{figure*}

\begin{figure*}[hbtp]
\begin{center}
\scalebox{1.0}{\includegraphics[width=18.0cm]{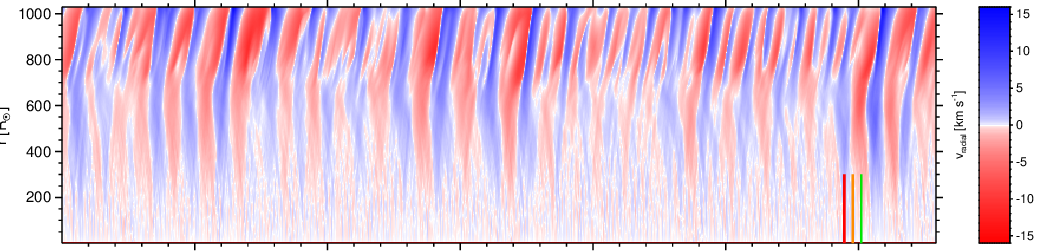}}
\scalebox{1.0}{\includegraphics[width=16.0cm]{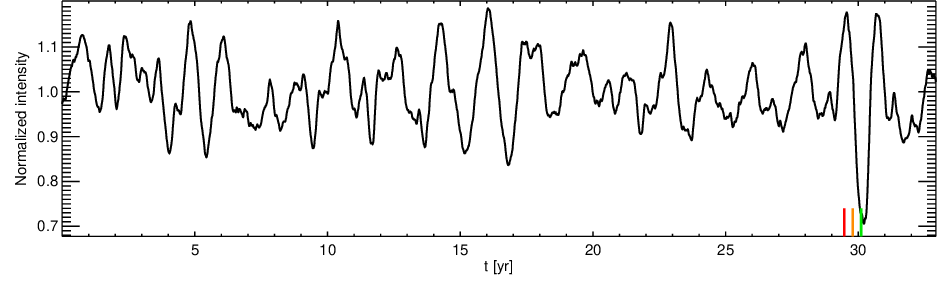}\hspace*{2cm}}
\end{center}
\caption{
  \label{f:st35gm04n048_QuOvertimeAndx}
  Temporal behavior of radial velocity and intensity
  of the 5\,$M_\sun$ model st35gm04n048.
  Top panel: Spherical averages of the
  radial velocity
  as a function of radial distance and time.
  Bottom panel: Normalized intensity (luminosity seen by an observer)
  corresponding to the view in Fig.\,\ref{f:st35gm04n048_0232501_QuSeq1}.
  The colored vertical lines (red, orange, green) at about $t$\,=\,30\,yr
  mark the instants selected for
  Figs.\,\ref{f:st35gm04n048_0232501_QuSeq1} and \ref{f:st35gm04n048_0232501_QuSeq3}.
}
\end{figure*}

It is obvious that the local and temporal variation in the emitted spectrum
is huge. This applies especially to the appearance and intensity
of the different molecular features.
For the data shown in Fig.\,\ref{f:spectra_BA},
there is a general trend that the absorption becomes stronger,
at later instants and in the area of the dark patch.
This is caused by lower temperatures in the region
where most of the emerging radiation is produced.
For the earliest epoch ($t$\,=\,6.781\,yr)
and the location not affected by the local cooling,
we find relatively weak bands of CO (around 2500\,nm)
and SiO (around 4000\,nm) in the infrared.
Also, the TiO features in the visual range are not strong,
as the corresponding gas temperatures are above 3500\,K.
The formation of the dark patch, on the other hand,
results in extremely deep absorption caused by TiO at shorter
and H$_2$O at longer (1500, 2000, 3000\,nm) wavelengths.
Such intense bands require material cooler than about 2600\,K in
the region where most of the radiation is produced.

For the last time step ($t$\,=\,7.289\,yr),
corresponding to the fully developed dark patch,
Fig.\,\ref{f:spectra_BA} also includes a continuum spectrum
computed without any line opacities.
This enables one to estimate the amount of molecular absorption.
One can see that in the region with local cooling,
TiO blocks almost all of the radiation in large parts of the visual range.
In combination with the weakening and shift of the continuum
due to the lower temperature, this results in a strong dimming
of the disc-integrated light
affecting especially the wavelength range around the V~filter,
where the fluxes may decrease by orders of magnitude
\citep[see also][]{Chiavassa2010A&A...511A..51C,Chiavassa2011A&A...535A..22C}.
The darkening gets steadily smaller, if one proceeds to longer wavelengths,
and also in the UV it is less pronounced.

\section{Enhanced convective downdraft and rebound}\label{s:NonstationaryConvection}

Local dimming events due to plumes caused by Rayleigh-Taylor instabilities,
as is discussed in Sect.\,\ref{s:Plumes},
are relatively common but usually not very strong.
A different mechanism is responsible for
the pronounced dimming event in model st35gm04n048
(see the sequence of bolometric images in Fig.\,\ref{f:st35gm04n048_0232501_QuSeq1},
with cross sections in Fig.\,\ref{f:st35gm04n048_0232501_QuSeq3}).
The cause is an exceptionally strong disturbance
in the global convective flow,
as is explained in the following.
It results in the single strong drop in the light curve
in Fig.\,\ref{f:st35gm04n048_QuOvertimeAndx}.
In principle, this scenario is often at work {in a weak form}.
However, we noticed only one very prominent occurrence in around a hundred models
\citep[more than those listed in][]{Ahmad2023A&A...669A..49A}.

Below the layers of small-scale surface granulation,
only a few large-scale cells dominate the convective flow.
This is seen in the left panels in Fig.\,\ref{f:st35gm04n045_0072609_QtauSeq},
where cool (dark red in the temperature row)
downdrafts (red in the velocity panel) delineate the large convective cells
in a snapshot of model st35gm04n045.
In the case of the dimming event in model st35gm04n048,
there are only two dominant cells with upward-moving material,
with a ring-shaped downdraft around the star.
In the first and third bottom panels
in Fig.\,\ref{f:st35gm04n048_0232501_QuSeq3},
the inward-flowing material
at the upper left and lower right of the star
are the signatures of the ring in these cross sections.

In addition to the rather persistent global downdraft ring,
there are other deep-reaching downdrafts,
which drift in the dominating convective flow,
finally merging with the downdraft ring
(see the two close-by downdrafts in the lower right of the star
as red, thin, wiggly lines pointing inward
in the bottom left panel of Fig.\,\ref{f:st35gm04n048_0232501_QuSeq3}).
Likely synchronized by the global pulsations
(see the top panel in Fig.\,\ref{f:st35gm04n048_QuOvertimeAndx}),
this seems to happen simultaneously at several locations on the ring.
{Before the merging, the flow of material from both sides
of each downdraft (i.e., four sides in total)
provides the mass transported inward by the two downdrafts.
For a short time during the merging process of the downdrafts,
the mass flow in the downdrafts is combined,
resulting in a single downdraft with twice the normal mass flux.
However, there is only material coming from two sides,
which is only sufficient for a single normal downdraft.}
This imbalance causes the local surface of the convection zone above the downflow
to move inward temporarily
(visible in the left column in Fig.\,\ref{f:st35gm04n048_0232501_QuSeq3}).

While the lateral mass flow still increases,
the downward mass flow of the downdraft after the merger reverts to the normal value
and the local surface of the convection zone rebounds
and reaches larger radii than usual
(see the center column in Fig.\,\ref{f:st35gm04n048_0232501_QuSeq3}).
As a consequence, material is pushed far above the convection zone
(see the right column in Fig.\,\ref{f:st35gm04n048_0232501_QuSeq3}),
essentially all around the downdraft ring.
However, there are strong local variations,
which explain why a major outflow occurs
only at the bottom right
in the right panels of Fig.\,\ref{f:st35gm04n048_0232501_QuSeq3}.

The overshooting material cools and dims, as is described for the
case of the plume formation via Rayleigh-Taylor instabilities
in Sect.\,\ref{s:DimmingScenario}.
However, in the current case, cool material is hovering all round the star
over the downdraft ring,
although with different amounts of material depending on the position.
A light curve, as in Fig.\,\ref{f:st35gm04n048_QuOvertimeAndx},
shows a very similar dimming event for the side opposite to the one chosen here.
For the other four orientations of the computational cube, however,
the minima are rather inconspicuous.
The view for the opposite side, also with large dark areas,
is shown in Fig.\,\ref{f:stXX_Int},
where a central vertical cut does not intersect a major dark patch,
and therefore does not show the effect in the cross sections
in Fig.\,\ref{f:st35gm04n048_0232501_QuSeq3}.

While dark patches on smaller scales are not individually visible
in the blurred images in Fig.\,\ref{f:st35gm04n048_0232501_QuSeq1},
close to the minimum of the light curve
the large dark area covers
a significant fraction of the stellar surface and remains visible
even after the blurring,
indicating that it would be detectable with current interferometers.
During the rebound,
a local shock wave is produced,
traveling outward in the lower right quadrant of the plots
shown in Fig.\,\ref{f:st35gm04n048_0232501_QuSeq3}.

\begin{figure*}[hbtp]
\begin{center}
\scalebox{1.0}{\includegraphics[width=18.0cm]{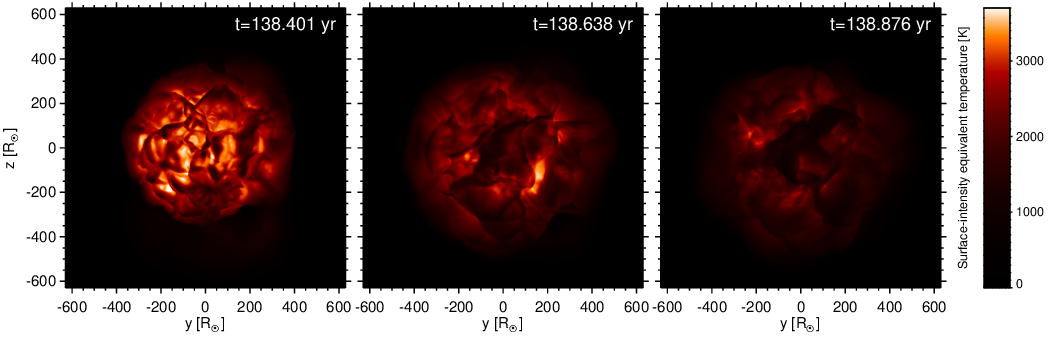}}

\scalebox{1.0}{\includegraphics[width=18.0cm]{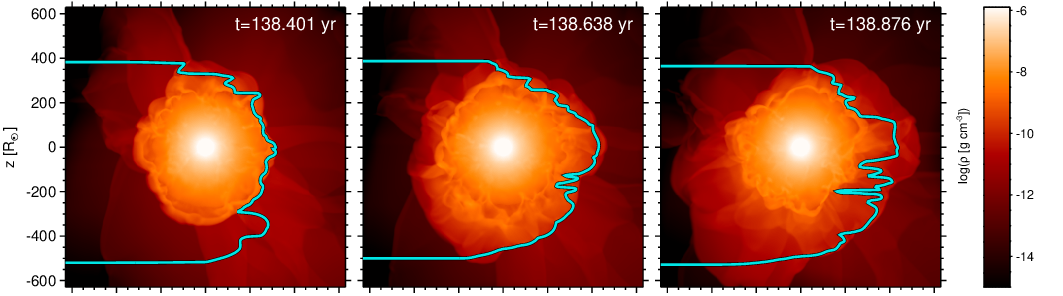}}

\scalebox{1.0}{\includegraphics[width=18.0cm]{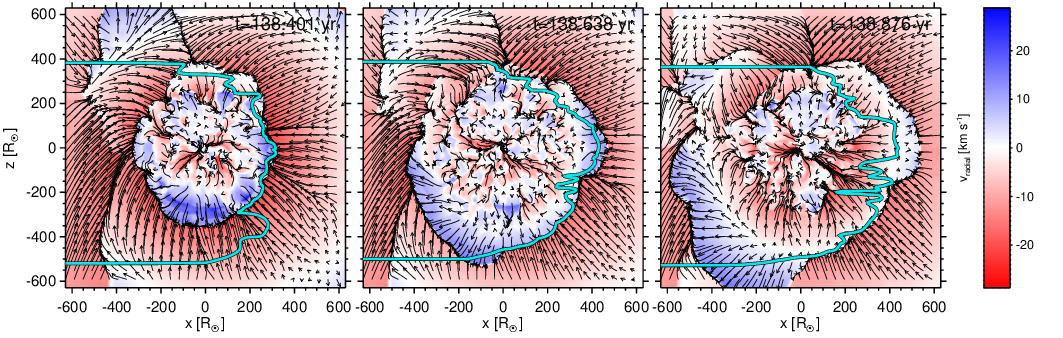}}
\end{center}
\caption{
  \label{f:st28gm05n056_0873500_QuSeq1}
  Time sequence of
  bolometric intensity in original resolution (top row)
  and a cross section though the logarithm of density (middle row),
  and the radial velocity
    with overlaid pseudo-streamlines (bottom row)
  of three snapshots of the 1\,$M_\sun$ model st28gm05n056.
  The cyan contour line indicates optical depth unity ($\tau_\mathrm{Ross}$\,=\,1)
  for an observer located to the right,
  corresponding to the viewing direction in the top-row images.
}
\end{figure*}

\begin{figure*}[hbtp]
\begin{center}
\scalebox{1.0}{\includegraphics[width=18.0cm]{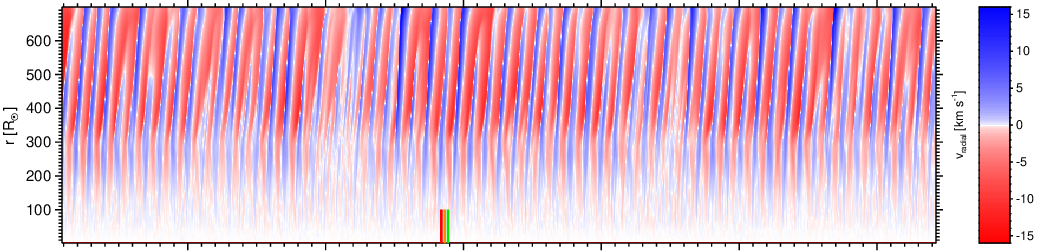}}
\scalebox{1.0}{\includegraphics[width=16.0cm]{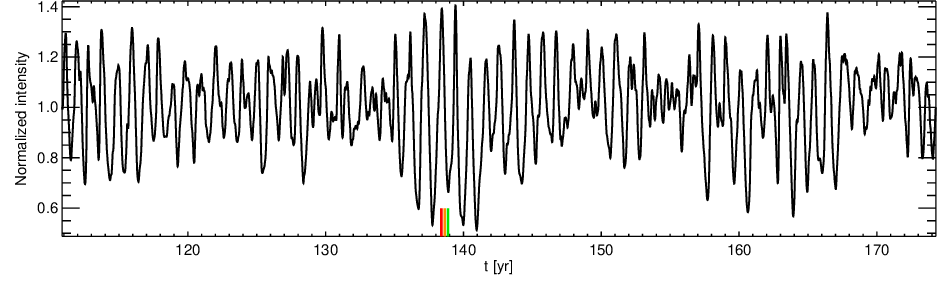}\hspace*{2cm}}
\end{center}
\caption{
  \label{f:st28gm05n056_QuOvertimeAndx}
  Temporal behavior of radial velocity and intensity
  of the the 1\,$M_\sun$ model st28gm05n056.
  Top panel: Spherical averages of
  radial velocity
  as a function of radial distance and time.
  Bottom panel: Normalized intensity (luminosity seen by an observer)
  corresponding to the view in Fig.\,\ref{f:st28gm05n056_0873500_QuSeq1}.
  The colored vertical lines (red, orange, green) at about $t$\,=\,138\,yr
  mark the instants selected for that figure.
  The timer for this model does not start at zero but at about $t$\,=\,111\,yr.
}
\end{figure*}

\section{Large-scale shock waves due to pulsations}\label{s:ShockDarkening}

Models of AGB stars with lower masses ($M_\star$\,=\,1 and 1.5\,$M_\sun$)
and lower effective temperatures ($T_\mathrm{eff}$\,$\lesssim$\,3000\,K)
than those presented in the previous sections
provide much more favorable conditions for the formation of molecules
in some layers of the atmosphere
during certain phases of the strong radial pulsations. Rather than localized patches on the surface,
these stars tend to periodically build up
a more or less spherical cocoon of dense and cool gas
(see a rather irregular example in
the right column of panels in Fig.\,\ref{f:st28gm05n056_0873500_QuSeq1}).
It may enshroud most of the star,
with -- possibly -- small regions that are transparent enough
to allow light from deeper, warmer layers to escape
(see the first two panels in Fig.\,\ref{f:stXX_Int}).
Similar time sequences have been shown by
\citet[][Figs.\,2 and 3]{Freytag2017A&A...600A.137F}.
Dark structures hovering over the brighter stellar surface were
already visible in the low-resolution (127$^3$ grid points) model used
by \citet[][Fig.\,1]{Freytag2008A&A...483..571F}.

During maximum light, only small dark areas exist close to the stellar surface
(left column in Fig.\,\ref{f:st28gm05n056_0873500_QuSeq1};
see also Fig.\,\ref{f:st28gm05n056_QuOvertimeAndx}).
During later phases of the pulsation cycle, a large patch develops,
and eventually the cool material almost completely
obscures the view onto the hotter stellar surface
(right column in Fig.\,\ref{f:st28gm05n056_0873500_QuSeq1}).

The cyan iso-surface-line at $\tau_\mathrm{Ross}$\,=\,1
in the middle and bottom rows in Fig.\,\ref{f:st28gm05n056_0873500_QuSeq1},
following the shock front outward at $z$\,$\approx$\,$-250\,R_\sun$,
indicates that the dark structure in the lower part of the star
(in the top row in Fig.\,\ref{f:st28gm05n056_0873500_QuSeq1})
is caused by the outward-moving shock.
However, the gas behind the shock front
is not sufficiently optically thick everywhere,
leaving small gaps where deeper and warmer material is visible.

Further density enhancements
caused by the azimuthal expansion and collision of post-shock material,
on the other hand,
can block the view toward the stellar surface very efficiently
(at $z \approx 0$ in the rightmost panels).
Compression of azimuthally moving material above the convection zone
can explain the structures at the boundaries of surface convection cells
in the very early low-resolution CO5BOLD models presented by
\cite{Freytag2002AN....323..213F}.
A dramatic collision of predominantly azimuthally moving material,
giving rise to a nearly radial and almost stationary shock,
is visible in the top region
of the two lower rows in Fig.\,\ref{f:st28gm05n056_0873500_QuSeq1}
(at $x$\,$\approx$\,0 and $z$\,$\approx$\,400 to 600\,$R_\sun$).
This shock occurs in optically thin regions,
so that the $\tau_\mathrm{Ross}$=1-iso-surface is hardly affected.

The signature of the lifting of the deep-surface material and
the typical triangular shape of the Rayleigh-Taylor instabilities,
described in Sect.\,\ref{s:Plumes},
can be seen at around $x$\,$\approx$\,0 and $z$\,$\approx$\,-250\,$R_\sun$
in the density and the velocity plots in Fig.\,\ref{f:st28gm05n056_0873500_QuSeq1}.
Convective motions and the rugged stellar surface induce some
structures in the extended atmospheric layers,
but are not as important for the occurrence of the dimming of the entire star
as in the scenario for hotter giants, described in Sect.\,\ref{s:DimmingScenario}.
This can explain why 1D (spherically symmetric) dynamical atmosphere and wind models 
for AGB stars give realistic spectra and photometric variations in the visual and 
near-infrared region, where pulsation effects on molecules dominate
\citep[see, e.g.,][]{Bladh2013A&A...553A..20B,
Bladh2015A&A...575A.105B,
Hoefner2016A&A...594A.108H,
Hoefner2022A&A...657A.109H}.

\section{Discussion}\label{s:Discussion}

\subsection{Ingredients of dimming scenarios and observational signatures}

All scenarios for the emergence of dark surface patches
due to clumps of cool molecular gas
presented in Sects.\,\ref{s:Plumes} to \ref{s:ShockDarkening}
involve convection, pulsations, and shock waves,
while the role and relative importance of these processes
depend on the stellar parameters.
Individually, the existence of these processes is well established by observations.
However, it is not easy to evaluate the contribution of the various mechanisms
for a certain dark surface structure or dimming event by observations.

Shocks traveling outward in the atmosphere
are produced during each pulsation cycle in all models.
Still, only in lower-mass AGB stars is the density of the gas in the post-shock regions sufficiently high to directly cause a darkening of the star,
as is described in Sect.\,\ref{s:ShockDarkening}.
During the minima of the light curve, almost the entire stellar surface
is obscured by cool, optically thick material.
Convection causes inhomogeneities in the cool post-shock gas
and might enhance the efficiency of the outward transport of material,
but it is not essential for the dimming mechanism as such.
In contrast, local dark areas in intermediate phases require the interaction
of the shock wave with convectively induced inhomogeneities in the lower atmosphere.

Plumes of cool gas due to Rayleigh-Taylor instabilities
(see Sect.\,\ref{s:Plumes})
occur very frequently in all models.
The extent of the resulting patches
ranges from the relatively small size of granules to the scales of global convection cells,
accompanied by a large variation in patch contrast.
However, to become significant for the overall appearance of the surface
and the light curve,
a particularly strong driving is needed,
which can be due to enhanced radial or non-radial acoustic pulsations
or fluctuations in the convective flow.

The scenario involving a convective collapse and rebound
can occasionally create very extended cool gas clouds causing prominent dark patches,
as is outlined in Sect.\,\ref{s:NonstationaryConvection}.
However, this seems to be a comparably rare event,
as in most cases only small, inconspicuous dark features are produced.

Because of the similarity of the produced dark patches (see Fig.\,\ref{f:stXX_Int})
that originate from different flow configurations well below the visible layers,
distinguishing between the scenarios observationally is not a trivial task.
For example, an outward-traveling shock wave is produced in all cases.
The timescales for the development of a patch are similar,
likely due to the coupling to stellar pulsations.
Both in the Rayleigh-Taylor and in the convective-rebound scenario,
cool matter moves outward,
while matter around it can fall back.
In the scenario with material moving outward behind a shock
(Sect.\,\ref{s:ShockDarkening}),
the matter just in front of the shock often has considerable infall velocities.

Large-scale Rayleigh-Taylor plumes occur predominantly over the upflow regions of 
global convection cells,
whereas material pushed up by a convective rebound is located above downdrafts.
The locations of post-shock clumps do not have a strong correlation with
convective up- or downflows.
In most cases,
we do not see the top layers of the convection zone directly.
The expected pattern of dark intergranular lanes surrounding bright granules
is often hidden by the material above.
In the model data, the pattern is apparent at optical depths
far beyond the reach of observations (see Fig.\,\ref{f:st35gm04n045_0072609_QtauSeq}).
At the surface, neither small-scale granules nor large-scale convective cells
have sufficient spatial brightness amplitudes
to explain observed large-scale high-contrast structures.
Therefore, the process in the convective envelope
responsible for the emergence of a dark patch
is not easily accessible to observations.
High-resolution surface images alone, as in Fig.\,\ref{f:stXX_Int}, are not sufficient
to distinguish between the scenarios.
Only time sequences of high-resolution images in combination with velocity information
derived from spectroscopic measurements may give us indirect hints to deduce
the mechanism responsible for the formation of a dark patch.

As convection plays a crucial role in creating the dark patches,
typical spatial scales of the dark patches and the underlying convective flow
are related.
It is just not directly possible to attribute, for example, a bright area
to a convective upflow region.

Additional physical conditions not included in the set of models presented here
might complicate the situation even more.
Rotation of the entire star or at least
the interaction of the surface layers with a rapidly rotating core
might induce a preferred axis for fluctuations at the poles.
The lowering of the effective gravitational acceleration at the equator
might facilitate the local levitation of gas.
Large-scale magnetic fields might inhibit convective motions
or channel waves and thereby modify the picture outlined above.
Sufficiently close planetary or stellar companions might play a role
for the dynamics, too.
While we presented a mechanism where the interaction of convection and pulsations alone
is able to produce puffs of cold molecular gas above the top of the convection zone,
we cannot exclude the contribution of other physical processes for a given case.

\subsection{Formation of dust}\label{s:DustFormation}

While the models analyzed in the current paper
do not consider dust physics,
they can serve to illustrate a mechanism for the formation of dust
that relies on 3D inhomogeneities
reaching sufficiently low temperatures and high gas density.

The pure shock scenario (see Sect.\,\ref{s:ShockDarkening})
works reliably to produce spherical dust shells and dust-driven winds
of cool, low-mass AGB stars, explored in detail with 1D models
\citep[see, e.g.,][]{Bladh2015A&A...575A.105B,
Bladh2019A&A...626A.100B,
Hoefner2016A&A...594A.108H,
Hoefner2022A&A...657A.109H}.
In 3D simulations,
additional azimuthal density fluctuations produce
clumpier, often arc-like dust clouds \citep[see][]{Hoefner2019A&A...623A.158H}.

In addition, as is demonstrated by \cite{Freytag2023A&A...669A.155F}
for models of an 1\,M$_\sun$ and an 1.5\,M$_\sun$ AGB star,
dust can form occasionally above a sufficiently large dark area
relatively close to the star.
Radiation pressure acting on the dust grains drives the dust cloud outward.
Further out, the shielding effect of the (likely dissolving)
dark patch diminishes, 
which might lead to heating and evaporation of the dust cloud.
Still, under the right conditions, some clouds survive and reach cool regions,
resulting in a clumpy, intermittent wind partly originating very close to the star.

The actual minimum temperature above a dark patch
(see Fig.\,\ref{f:st35gm04n045_1DRay})
is even lower than the bolometric temperature
in Fig.\,\ref{f:st35gm04n045_0069300_QuSeq2}.
With about 1500\,K, it comes very close to the condensation temperature of corundum
\citep[see][]{Hoefner2016A&A...594A.108H,
Hoefner2019A&A...623A.158H}.
Corundum grains may serve as cores on which a mantle of silicates can form later.

Due to its optical properties,
corundum efficiently re-emits absorbed stellar radiation,
resulting in a lower temperature of the dust grains than that of the surrounding gas.
Therefore, grains can sustain slightly higher (gas) temperatures
than they need for their formation.
That can help a dust cloud to survive
when the dark surface patch vanishes
and the dust has not yet reached a distance
where the temperatures are always below the condensation temperature.

All models presented in the current paper use
gray Rosseland opacities.
\cite{Hoefner2019A&A...623A.158H} have shown that using opacity tables
that take the frequency dependence of the opacities into account
gives much more favorable conditions for the formation of dust.


\subsection{Pressure contributions}\label{s:RadAndDynPressure}

\begin{figure}[hbtp]
\begin{center}
\includegraphics[width=8.8cm]{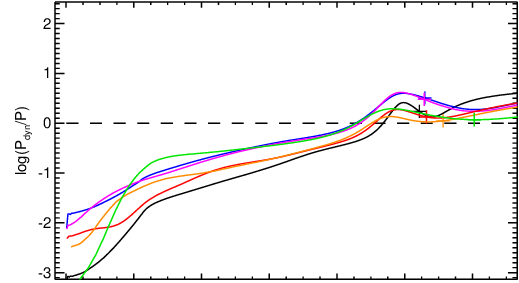}\vspace{-0.05cm}
\includegraphics[width=8.8cm]{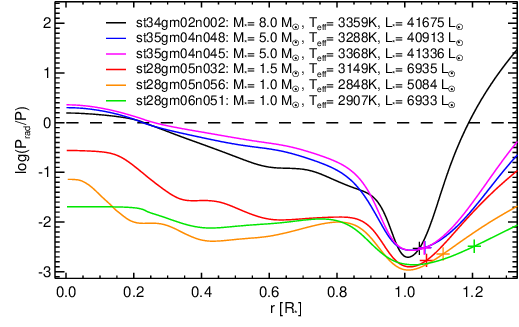}\vspace{-0.2cm}
\end{center}
\caption{
  \label{f:RadAndDynPressure}
Quantities averaged over spherical shells and time
versus relative stellar radius,
for the models in Fig.\,\ref{f:stXX_Int} and Tab.\,\ref{t:ModelParam}.
Top panel:
Ratio of radial dynamical pressure to gas pressure.
Bottom panel:
Averaged ratio of radiation pressure (in optically thick approximation) to gas pressure.
The horizonal dashed lines mark pressure balance $P_\mathrm{rad,dyn}=P$, respectively.
The plus signs indicate the approximate positions of Rosseland depth unity.
}
\end{figure}

The relative importance of gas pressure, dynamical pressure, and radiation pressure
can be estimated from Fig.\,\ref{f:RadAndDynPressure}.
The top panel shows the averaged ratio of radial dynamical pressure and gas pressure,
thereby condensing the complex processes described in the previous sections into
simple one-dimensional curves.
The dominance of the dynamical pressure over the gas pressure in the outer layers,
starting even below the surface,
is an indication of the importance of the various dynamical processes for the
structure of these layers.

The radiation pressure
(bottom panel of Fig.\,\ref{f:RadAndDynPressure})
is not included in the simulations but estimated from
$P_\mathsf{rad}=(4 \sigma_\mathsf{SB})/(3 c_\mathsf{light}) T^4$,
which is strictly valid only in the optically thick regime.
As is described in Sect.\,\ref{s:DustFormation},
it can become important in the outer atmospheric layers,
if the opacities are large enough so that gas or dust absorb or scatter
a sufficient amount of photons.
The importance of radiation pressure in the inner region
and its neglect in the simulations,
together with other approximations (described Sect.\,\ref{s:GeneralSetup})
could be responsible for shortcomings of the models in these layers.
However, close to the stellar surface, the radiation pressure is small
compared to the gas pressure,
with the dynamical pressure dominating both of them.
In these layers,
temperatures are low while gas densities and Mach numbers are high, so that the
dynamical pressure is up to three orders of magnitude higher than the radiation
pressure.
Therefore, we expect the inclusion of radiation pressure to have a minor effect
on the processes described in the previous sections,
while it is crucial for the formation of a dust-driven wind (Sect.\,\ref{s:DustFormation}),
and important for the stellar interior.

\subsection{The chromosphere}\label{s:ChromoSphere}

The atmospheres of our models are filled with
a dynamical network of acoustic shock fronts,
in which the gas is compressed and heated but cools down rapidly
(the green, hot filaments in the blue, cold outer atmosphere in the temperature plots
in Figs.\,\ref{f:st35gm04n045_0069300_QuSeq3} and \ref{f:st35gm04n048_0232501_QuSeq3}).
We assume that the radiative cooling from the shocks is responsible
for at least a part of the chromospheric emission
according to the picture
of an acoustically heated chromosphere,
already discussed, for instance, by
\cite{Vitense1953ZA.....32..135V}.
\cite{Carlsson1992ApJ...397L..59C} presented detailed 1D RHD models
of a purely acoustically heated solar chromosphere.
\cite{Wedemeyer2004A&A...414.1121W} and \cite{Leenaarts2012ApJ...749..136L}
performed multidimensional local simulations of the solar chromosphere,
and \cite{Wedemeyer2017A&A...606A..26W} presented similar simulations of the
chromospheres of red-giants stars.
Although our simulations show the ubiquitous shocks,
they have neither the high numerical resolution
nor the sophisticated non-LTE radiation-transport calculations
necessary to model the radiative properties of the shocks in detail.

In contrast to waves where the energy is contained within the magnetic field,
the transport of energy by acoustic waves only works efficiently as long as the
gas density is sufficiently high.
However, an extended region with high gas densities
due to molecular plumes or a stellar wind
would allow for significant chromospheric emission by acoustic heating even further out.
A wind
(driven by radiation pressure on dust formed in cold post-shock regions)
might therefore be a vital prerequisite
to create the conditions for a chromosphere
(a network of hot shocks at sufficiently high gas densities)
in AGB and RSG stars.
Hot chromospheric gas (in the shock fronts)
and cold gas able to form molecules or even dust (in the inter-shock regions)
cannot exist at the same point in space and time
but can easily occur at the same distance from the star
or in the same line of sight.

\subsection{Differences between solar and RSG surface conditions}\label{s:SolarvsRSGSurface}

The about five orders of magnitude smaller surface gravity of RSGs compared to
the Sun leads to a correspondingly larger pressure scale height
(ignoring comparatively small differences in temperature and mean molecular weight).
Even the ratio of pressure scale height to stellar radius 
(essentially given by $(R_*/R_\sun)/(M_*/M_\sun)$)
is significantly larger.

That results in not only much bigger but also fewer surface granules
\citep{Schwarzschild1975ApJ...195..137S},
on the one hand.
The slower density increase inward,
on the other hand,
causes a slower decrease -- or even an increase -- in convective velocities
\citep[see, e.g.,][]{Goldberg2022ApJ...929..156G}.
Thereby, the deeper layers gain importance and are able to contribute
to the driving of stronger large-scale (global) convection cells
becoming important for the surface flows
\citep[see, e.g.,][]{Stothers1971A&A....10..290S}.
A third consequence is the lowered escape velocity
facilitating the formation of a stellar wind.

While solar-type stars show a spectrum of millions of small-amplitude p-modes,
RSGs pulsate in a single p-mode or at most very few modes,
albeit with much larger amplitudes.
Therefore, the standing waves in the interior immediately are transformed into
traveling shocks when reaching the thin and cool atmospheric layers.
In the Sun, they pass the relatively quiet solar photosphere
as traveling small-amplitude waves,
before reaching (and possibly powering) the chromosphere as shock waves
\citep{Wedemeyer2004A&A...414.1121W}.

The much smaller gravity (and therefore density)
and only somewhat smaller effective temperature (and therefore energy flux)
in RSGs
cause larger convective velocities and -- in particular -- Mach numbers.
The resulting much more significant dynamical pressure
translates into generally much more violent dynamics.

The lower effective temperature, and therefore lower typical temperatures in
the atmospheres of RSGs,
need smaller negative temperature fluctuations to allow for the formation of
significant amounts of molecules like TiO or water
and a corresponding significant increase
in the opacities (see Fig.\,\ref{f:st35gm04n045_opa}).
Even temperatures low enough to allow dust to form might occur occasionally.
In combination,
the larger scales,
higher Mach numbers,
and very different atmospheric opacities
render the surface appearance of RSGs completely dissimilar
to the well-known pattern of solar granulation.

\subsection{The 2019/2020 dimming event of Betelgeuse}\label{s:BetelgeuseDimming}

We analyzed representative CO5BOLD models of cool giant stars
between 1 and 8\,$M_\sun$
and identified three processes that cause -- at first glance similar --
dark structures in the stellar atmospheres
(see Fig.\,\ref{f:stXX_Int}).
Large-scale convection cells
interacting with
radial and non-radial pulsation modes
and pushing large amounts of gas into layers just above the convection zone
are the underlying cause for the structures.
While these commonly occur in nearly all of the models,
the details depend on the stellar parameters
that control convection and pulsations.
The effective temperature is a measure for the energy flux to be
transported through the atmosphere but also influences the opacities
(see Fig.\,\ref{f:st35gm04n045_opa}).
Surface gravity controls the size of small-scale granules but
also the gas density and therefore, together with the effective temperature,
convective velocities.
In addition, the pulsation properties (excited modes, periods, amplitudes)
depend on stellar parameters.

At present, CO5BOLD models of higher-mass RSGs 
suffer from an insufficient description of the stellar interior
at higher temperatures
(likely the missing radiation pressure,
possibly the missing self-gravity,
and a contribution from the damping ``inner boundary condition'').
This effectively reduces the amplitudes of pulsations and large-scale convective flows,
presumably causing the too compact atmosphere
\citep{ArroyoTorres2015A&A...575A..50A}.
Therefore, the application of our models
to the case of Betelgeuse (with $M_*$\,$>$\,10\,$M_\sun$)
needs some extrapolation beyond the results for the lower-mass stars.
The darkening due to cool post-shock gas
(see Sect.\,\ref{s:ShockDarkening})
requires large pulsation amplitudes and low effective temperatures, which are
mostly found in lower-mass AGB stars.
However, the other two scenarios involving Rayleigh-Taylor plumes
(see Sect.\,\ref{s:Plumes})
or a convective rebound
(see Sect.\,\ref{s:NonstationaryConvection})
are candidates to explain the process causing
the Great Dimming of Betelgeuse.

An orbiting dust cloud passing at a distance in front of the star
seems to be ruled out by the sequence of VLTI images taken by
\cite{Montarges2021Natur.594..365M},
that shows the darkening coming and going in the same region
of the stellar disk.
In some media, artistic renderings of the dark structure seem to be inspired
by sunspots -- regions on the solar surface where convective energy transport
is inhibited.
However, the most discussed process involves darkening by cool gas or even dust
somehow lifted above the bright stellar surface
\citep[e.g.,][]{Guinan2020ATel13410....1G,
Wheeler2023A&G....64.3.11W}.
Our simulations underpin this scenario, in particular supporting
the picture of large-scale ``molecular plumes'' observed by
\cite{Kervella2018A&A...609A..67K}, well before the Great Dimming.

While stationary convection alone causes some overshoot of material into the
stable layers above,
it is the interaction with pulsations that leads to (locally) strongly enhanced
ejections of material sufficient to obscure part of the bright stellar surface.
The enhanced dimming due to a convective rebound seen in
Fig.\,\ref{f:st35gm04n048_QuOvertimeAndx} is in sync with the stellar pulsation cycle,
in agreement to what has been observed for Betelgeuse
\citep[see, e.g.,][]{Guinan2019ATel13365....1G,
Lloyd2020BAAVC.184...22L,
Jadlovsky2023NewA...9901962J}.

The dimming of Betelgeuse was accompanied only by a moderate reduction of the
spectroscopically determined effective temperature,
not sufficient to explain the amount of brightness reduction,
suggesting either an unrealistic shrinking of the star or a local obscuration
\citep[see, e.g.,][]{Mittag2023A&A...669A...9M}.
\cite{Harper2020ApJ...905...34H} proposed a two-component model
with the normal stellar surface of about 3650\,K
and a dark component with $T$$\leq$3400\,K to explain a number of observations
without the need for any dust.
The cuts through our model (in Fig.\,\ref{f:st35gm04n045_0069300_QuSeq2})
show local effective temperatures going down to 2000\,K,
suggesting a three-component model with
the normal stellar surface,
the borders of the dark patches with a slightly reduced temperature,
and the core regions of the patches with a very {low} temperature and surface brightness.
As a consequence,
the latter will only show a negligible contribution to the overall spectrum.
(see Sect.\,\ref{s:Spectra}).

In the RHD simulations, large-amplitude pulsations
are essentially standing waves in the stellar interior,
accompanied by outward-traveling shock waves in the atmosphere
\citep[see Figs.\,\ref{f:st35gm04n045_QuOvertimeAndx}
and \ref{f:st35gm04n048_QuOvertimeAndx}
and a detailed analysis in][]{Ahmad2023A&A...669A..49A}.
\cite{Kravchenko2021A&A...650L..17K} detected two shock fronts
in the atmosphere of Betelgeuse, just before the dimming event.
Observations with the HST of the chromospheric activity of Betelgeuse during the
dimming event by \cite{Dupree2020ApJ...899...68D} supported the scenario
of a chromosphere heated by acoustic shocks
\citep[see, for instance,][]{Carlsson1992ApJ...397L..59C,
Wedemeyer2017A&A...606A..26W}.

It has been successfully tried to quantitatively model the dimming by an obscuring dust cloud
\citep[e.g.,][]{Montarges2021Natur.594..365M,
Cannon2023A&A...675A..46C}.
However, due to the absence of any detected unambiguous dust features,
an explanation just with cooler gas on or just above the surface
is equally possible
\citep{Dharmawardena2020ApJ...897L...9D,
Harper2020ApJ...905...34H,
Cannon2023A&A...675A..46C}.
The RHD models we present in the current paper lack any description of dust,
and the dimming is solely due to the ejection and cooling of gas.
However, the process of dust forming just above a cool patch of gas close to
the surface of an 1.5\,$M_\sun$ AGB star,
as is presented by \cite{Freytag2023A&A...669A.155F},
might even work occasionally under the more extreme conditions of RSGs
and play a role for the formation of the wind responsible for the observed
dust around Betelgeuse
\citep[e.g.,][]{Haubois2023A&A...679A...8H}.

\section{Conclusions}\label{s:Conclusions}

We performed ``star-in-a-box'' simulations
with the RHD code CO5BOLD
to produce a number of global models of AGB and RSG stars,
covering the pulsating outer convection zone
and the shock-filled inner atmosphere.
From those, we selected a sample set,
with current stellar masses between 1 and 8\,$M_\sun$,
to demonstrate three different physical mechanisms
that are able to lift dense clouds of material above the top of the convection zone.
This rapidly cooling gas appears as localized dark patches
when seen against the surface of the stellar models.

We picked a cool 1\,$M_\sun$ model as a typical example of a low-mass AGB star
to show how the process of material transport in the wake of a strong shock,
that is well known from 1D (spherically symmetric) simulations,
behaves in 3D geometry.
The relatively hot stellar surface (at light curve maxima)
and the enshrouding cocoon of cooler molecular gas (at light curve minima)
do not show very pronounced large-scale asymmetries.
However, phases in between tend to reveal large dark patches,
caused by inhomogeneities in the lower atmosphere,
induced by convective motions and possibly non-radial acoustic modes.

In addition, we analyzed the formation of a dark patch
on the surface of a slightly warmer 5\,$M_\sun$ model.
In this case,
the patch is the visible signature of cool, dense gas lifted above
the surface of the convection zone by plumes
forming from Rayleigh-Taylor instabilities in the flow.
They arise from the
combined updraft of a global-scale convection cell
and an acoustic wave traveling outward through the surface layers.
This is a common process in all our global models,
with a wide range in size and contrast of the dark patches,
depending on the amplitude and geometry of the convective and
pulsational velocity fields,
that are a function of the stellar parameters.

Likewise, another scenario happens frequently on small scales.
It involves large-amplitude convective fluctuations that first lead to enhanced flows in deep downdrafts and
that then rebound
and send material outward into the atmosphere
above the convective downdrafts.
However, we only found one very strong example, in another 5\,$M_\sun$ model,
where the process is able to lift up sufficient amounts of material,
so that the bolometric light curve shows a prominent dimming event.

In all cases, the elevated material cools sufficiently
and temperatures become so low
that large amounts of molecules like TiO or water start to form,
increasing the gas opacities drastically.
In the lower-mass AGB-star models of \cite{Freytag2023A&A...669A.155F},
silicate dust can form in small pockets
in the coolest top-part of the molecular plumes.
The dust clouds are then accelerated by radiation pressure and,
under favorable conditions,
some of the clouds can reach cool layers further out,
where dust grains are stable,
thus contributing to a nonstationary, non-isotropic stellar wind.

We assume that the basic mechanisms causing
shocks, Rayleigh-Taylor plumes, or convective rebounds
will also work in the slightly hotter atmospheres of massive RSGs; that is, under somewhat more extreme conditions than the ones found in our sample of stars.
Via these mechanisms,
the fundamental processes convection and pulsations,
widely assumed to be responsible for the Great Dimming of Betelgeuse
\citep[see][]{Wheeler2023A&G....64.3.11W},
can actually affect the photosphere sufficiently
to form a large-scale dark patch of cold molecular gas.
We speculate that even in an RSG,
at the top of a particularly strong and cold molecular plume,
dust species with high condensation temperature (for example Al$_2$O$_3$)
might form -- very close to the star.
In this way, dust grains could form as a consequence of a local dark patch,
but would not be the primary cause of a dimming event, as was observed for Betelgeuse.

\begin{acknowledgements}

This work is part of a project that has received funding from 
the European Research Council (ERC) 
under the European Union’s Horizon 2020 research and innovation programme 
(Grant agreement No.~883867, project EXWINGS) 
and the Swedish Research Council ({\it Vetenskapsr{\aa}det}, grant number 2019-04059). 
The computations were enabled by resources provided by
the Swedish National Infrastructure for Computing (SNIC)
and the National Academic Infrastructure for Supercomputing in Sweden (NAISS),
partially funded by the Swedish Research Council through grant agreements
no.~2018-05973
and no.~2022-06725.

\end{acknowledgements}

\bibliographystyle{aa}    
\bibliography{aa_redsg}


\end{document}